\documentclass[fleqn,usenatbib]{mnras}

\usepackage{newtxtext,newtxmath}
\usepackage[T1]{fontenc}
\usepackage{threeparttable}
\usepackage{caption}
\DeclareRobustCommand{\VAN}[3]{#2}
\let\VANthebibliography\thebibliography
\def\thebibliography{\DeclareRobustCommand{\VAN}[3]{##3}\VANthebibliography}

\usepackage{graphicx}	% Including figure files
\usepackage{amsmath}	% Advanced maths commands
\usepackage{overpic}
\usepackage{ulem}
\title[``Extreme'' GRG simulations]{Simulating megaparsec-scale jets of radio galaxies: Magneto-hydrodynamics of jets reaching 5 Mpc.}

\author[G. Giri et al.]{
Gourab Giri,$^{1}$\thanks{E-mail: g.giri@ira.inaf.it}
Dario Borgogno,$^{2}$
Marco Tavani,$^{2}$
Andrea Mignone,$^{3}$
Prateek Sharma,$^{4}$
Claudio Gheller,$^{1}$ 
\newauthor{Valerio Vittorini,$^{2}$
Paul J. Wiita,$^{5}$
Bernie Fanaroff,$^{6}$
Alessio Suriano,$^{3}$
D. J. Saikia,$^{7,8}$
Gianfranco Brunetti$^{1}$}
\\
$^{1}$Istituto Nazionale di Astrofisica (INAF) – Istituto di Radioastronomia (IRA), via Gobetti 101, 40129 Bologna, Italy\\
$^{2}$Istituto Nazionale di Astrofisica (INAF) - Istituto di Astrofisica e Planetologia Spaziali (IAPS), via del Fosso del Cavaliere 100, 00133 Roma, Italy\\
$^{3}$Dipartimento di Fisica, Università degli Studi di Torino, Via Pietro Giuria 1, I-10125 Torino, Italy\\
$^{4}$Department of Physics, Indian Institute of Science, Bangalore 560012, India\\
$^{5}$Department of Physics, The College of New Jersey, 2000 Pennington Road, Ewing, NJ 08628-0718, USA\\
$^{6}$South African Radio Astronomy Observatory, 2 Fir Street, Black River Park, Observatory, Cape Town 7925, South Africa\\
$^{7}$ Fakultat f\"ur Physik, Universit\"at Bielefeld, Postfach 100131, D-33501 Bielefeld, Germany \\
$^{8}$ Assam Don Bosco University, Guwahati 781017, Assam, India \\
}

\date{Accepted XXX. Received YYY; in original form ZZZ}

\pubyear{\the\year{}}

\begin{document}
\label{firstpage}
\pagerange{\pageref{firstpage}--\pageref{lastpage}}
\maketitle

\begin{abstract}
Extragalactic jets have long prompted the question of how far relativistic outflows can extend, with some radio sources reaching 5–7 Mpc in length.  These great extents motivate investigations into their ages, propagation dynamics, stability, and impact on the environment. 
We perform 3D high-resolution numerical simulations of two jet configurations involving continuous injection at different powers propagating in low-density regions of the cosmos (static and laminar), investigating the conditions for jet collimation versus disruption at extreme scales. 
We show that the combined effects of higher jet thrust (enhanced kinetic power), improved collimation (suppression of transverse distortions), and magnetic stabilization (strengthened poloidal field) can sustain a laterally confined flow, enabling such a jet to reach 5 Mpc in just 15 Myr (injecting a total energy of $2.3 \times 10^{61}$ erg into the environment). 
In contrast, a jet lacking these conditions dissipates more rapidly, forming lobe-like morphologies and reaching only $\sim 3$ Mpc over $\sim35$ Myr (injecting total energy of $8.1 \times 10^{60}$ erg). 
Pinch and kink MHD instabilities are identified as the primary drivers of transverse distortions; their suppression allows the persistence of a fast spine alongside a slower, dissipative head (location of maximum environmental interaction). 
We find that the jet-head propagation shows two regimes: one with speed $\sim0.5 c$; the other with speed from $\sim 0.2 c$ to $\sim 0.05 c$. 
We consider a proxy of synchrotron emission and find that radiation is concentrated in regions of enhanced compression and magnetic amplification, primarily near the first recollimation shock (producing a bright radio spot) and at the jet-head interaction zone (producing the radio termination lobe). 
Such jets facilitate the transport of substantial energy and magnetic flux into underdense cosmic regions.
\end{abstract}

\begin{keywords}
galaxies: groups: general -- galaxies: jets -- methods: numerical -- (magnetohydrodynamics) MHD -- instabilities
\end{keywords}

%%%%%%%%%%%%%%%%%%%%%%%%%%%%%%%%%%%%%%%%%%%%%%%%%%

%%%%%%%%%%%%%%%%% BODY OF PAPER %%%%%%%%%%%%%%%%%%

\section{Introduction}

Since the early studies of extragalactic radio sources \citep[e.g.,][]{Jennison1953}, one of the central questions has been how far the astrophysical flows feeding the extended lobes can propagate before experiencing complete attenuation. 
Prior to the discovery of 3C~236 \citep{Willis1974}, whose reported linear size was 5.7~Mpc (revised to $\sim 4.1$~Mpc under a present-day cosmology with $H_0 = 70\ \text{km s}^{-1}\ \text{Mpc}^{-1}$), it was generally believed that the maximum extent of Active Galactic Nuclei (AGN) jets was limited to $\sim 1$~Mpc. 
For nearly three decades, 3C~236 remained the record holder until the identification of J1420$-$0545, with a projected jet size of 4.69~Mpc \citep{Machalski2008}. 
In the recent past, however, substantial advances in radio interferometry—both in sensitivity and resolution \citep[e.g., LOFAR, MeerKAT;][]{vanHaarlem2013,Jonas2016}—have enabled the detection of even larger systems, culminating in the discovery of giant radio galaxies with total extents in the range of 5–7~Mpc \citep{Oei2022_5Mpc,Oei2024_7Mpc}.

Giant radio galaxies (GRGs) are generally defined as extragalactic jets with projected sizes exceeding $0.7$~Mpc \citep{Dabhade2023}. 
While this threshold lacks a strong physical justification, recent numerical simulations by \citet{Giri2025_GRGSim} suggest a mild shift in the dynamical behavior of the lobes at these scales manifested in the (rate of change of) lobe expansion speed and internal pressure evolution around these scales. 
Such results provide theoretical motivation for distinguishing GRGs from their smaller counterparts, often referred to as standard radio galaxies (SRGs). However, this trend remains preliminary and requires validation across a broader region of parameter space.
The advent of modern interferometric facilities has dramatically accelerated the discovery of GRGs. 
The known population has expanded from $\sim 750$ sources in 2021 \citep[e.g.,][]{Dabhade2020_LOTSS,Dabhade2020_VLASS} to $\sim 2050$ sources in 2023 \citep{Oei2023_lenghDist}, and most recently to $\sim 11500$ GRGs \citep{Mostert2024}. 
This rapid growth has been enabled by advances in both data analysis techniques and machine learning algorithms, leading to sky densities of GRGs that now rival those of luminous non-giant radio galaxies. 

Along with the rapid growth of the number of GRGs the longstanding question of how far extragalactic jets can propagate remains unresolved. 
Until recently, GRGs with projected extents of $\gtrsim 3$~Mpc were exceedingly rare \citep{Andernach2021}, and those exceeding 5~Mpc were considered `extreme' cases. 
However, with the surge in GRG discoveries, the sample of such large-scale sources has expanded substantially. 
For example, \citet{Andernach2025} report 142 (candidate) GRGs exceeding 3~Mpc in projected size, while \citet{Hardcastle2023} identify 13 (candidate) sources larger than 4~Mpc. 
The extent-wise upper bound of these jetted-sources continues to be pushed, exemplified by the recent detection of the $\sim 7$~Mpc long `Porphyron' \citep{Oei2024_7Mpc}, representing the largest known extragalactic source. 

In this context, it has become increasingly meaningful to investigate the formation processes of GRGs from a numerical perspective, particularly as the inferred evolutionary mechanisms of extreme cases derived from observations appear to challenge the standard framework of jet propagation physics. 
While it has long been predicted that GRGs preferentially evolve in low-density environments such as galactic filaments or at the peripheries of dense cosmic structures \citep{Subrahmanyan2008,Malarecki2013}, recent findings by \citet{Sankhyayan2024} demonstrate that such a scenario cannot fully account for a substantial GRG population. 
Specifically, a significant fraction of GRGs are now observed in association with the brightest cluster galaxies (BCGs) within superclusters \citep{Sankhyayan2024}. 
Moreover, even radio galaxies exceeding $\sim 3$~Mpc have been found to evolve within clusters and rich groups, with many serving as BCGs themselves \citep{Andernach2025,Charlton2025}. 
Furthermore, while the majority of GRGs exhibit FR-II morphology—consistent with the need for powerful jets to overcome environmental resistance on large scales \citep{Dabhade2020_LOTSS,Simonte2024}—the existence of FR-I type low-powered GRGs extending up to 5~Mpc \citep{Oei2022_5Mpc} raises important questions. 
Adding to this puzzle, giant radio quasars (GRQs), despite their typically higher jet powers \citep{Mahato2022}, struggle to achieve such extreme sizes, with the largest known GRQ reaching 4.45~Mpc \citep{Coziol2017}. 
However, as per the unification scheme, quasars are expected to be seen at smaller angles to the line of sight than radio galaxies. 
For an inclination angle of $\sim$45$^\circ$, the dividing line between radio galaxies and quasars suggested by \citet{Barthel1989ApJ...336..606B}, a projected size of 4.45~Mpc corresponds to an intrinsic size of $\sim$6.3 Mpc.

Some models interpret GRGs as the oldest jets in the cosmos \citep{Hardcastle2019}. 
However, the long-term persistence of synchrotron emission raises a significant challenge to this scenario. 
For instance, non-thermal particles in equipartition, having a median $B \sim 5\ \mu$G field with Lorentz factor $\gamma_e \approx 10^4$ \citep{Croston2005,Giri2022_XRG,Dabhade2023} have a synchrotron cooling time of about $\sim 70$~Myr, which stands at odds with the Gyr-scale ages that have been inferred for examples like `Alcyoneus' and `Porphyron' \citep{Oei2022_5Mpc,Oei2024_7Mpc}. 
In addition, given the large sizes of GRGs, adiabatic losses and inverse-Compton scattering off cosmic microwave background photons become important \citep{Ishwara-chandra1999,O'dea1997,Schoenmakers2000,Konar2004,Giri2025_Emission}. Several studies also predict lower evolutionary ages for GRGs, comparable to or smaller than those of SRGs \citep{Konar2004,Sebastian2018}. 
An alternative scenario associates GRGs with restarted double-double radio galaxies (DDRG-GRGs), where a renewed jet propagates faster through cavities created by earlier outbursts, potentially explaining rapid growth \citep{Subrahmanyan1996}. 
While this mechanism may account for a subset of GRGs, the fraction of DDRG-GRGs remains low \citep{Cotton2020,Dabhade2025,Andernach2025}, and evidence of jet restarts in many `extreme' GRGs is not immediately apparent. 
Recent magneto-hydrodynamical simulations \citep{Duan2025,Giri2025_GRGSim,Giri2025_Emission} have begun to address some of these issues, demonstrating the evolution of GRGs reaching $\geq 1$~Mpc.

It has now been recognized that understanding the relative role of individual growth factors enabling GRGs to reach extreme scales requires focused investigation of the most extended examples. 
A central question concerns how the relativistic jet spine can preserve its collimation over Gyr timescales despite the expected disruption from magneto-hydrodynamical instabilities \citep{Oei2024_7Mpc} and entrainment of ambient matter during jet–lobe evolution \citep{Rossi2024}. 
The severity of this problem is underscored by alternative models proposed for sources such as Porphyron ($\sim 7$ Mpc), invoking scenarios such as $\gamma$-ray photon beam propagation from the host galaxy's nuclei rather than conventional jet-driven outflows \citep{Nernonov2025}. 
While such interpretations offer intriguing perspectives, they remain difficult to reconcile with the growing population of GRGs reaching $\sim 5$~Mpc that display classical jet morphologies and dynamics.

In this context, the present study addresses the pressing need for numerical verification of the evolution of `extreme' GRGs. 
By employing magneto-hydrodynamical simulations capable of producing oppositely directed jet pairs that extend to $\sim 5$~Mpc, we directly test the robustness of the standard jet paradigm under conditions relevant to the largest known radio galaxies. 
This approach not only constrains the physical mechanisms that allow jets to remain collimated and dynamically stable over spatial scales of several Mpc but also provides an avenue to reconcile theoretical expectations with the observed population of extreme GRGs in the pre-Square Kilometre Array\footnote{\texttt{SKAO:} \url{https://www.skao.int/en}} era. 

Section~\ref{Sec:Simulation setup} outlines the numerical framework employed in this study. The temporal evolution of these systems is analyzed in Section~\ref{Sec:Morphological and Temporal Evolution} and the intricate dynamics of jet propagation are examined in Section~\ref{Sec:Dynamics of Jet Propagation}. The potential feedback of these jets on the ambient cosmic environment is investigated in Section~\ref{Sec:Jet -- environment interconnection}. Finally, the principal conclusions are summarized in Section~\ref{Sec:Conclusions}.

%%%%%%%%%%%%%%%%%%%%%%%%%%%%%%%%%%%%%%%%%%%%%%%%%%%%%%%%%%%%%%
\section{Simulation setup}\label{Sec:Simulation setup}
%
%
%%%%%%%%%%%%%%%%%%%%%%%%%%%%%%%%%%%%%%%%%%%%%%%%%%%%%%%%%%%%%%

We performed our simulations using \texttt{gPLUTO}\footnote{\texttt{gPLUTO:} \url{https://plutocode.ph.unito.it/pluto-gpu.html}} \citep{Rossazza2025}, the GPU-accelerated successor of the widely used astrophysical code \textsc{PLUTO} \citep{Mignone2007}. 
\texttt{gPLUTO} is a full C++ rewrite of \textsc{PLUTO}, designed to extend its capabilities to exascale high-performance computing environments by supporting both CPU and GPU parallel computations. 
It preserves the core philosophy and numerical schemes of its predecessor, while integrating optimized GPU acceleration through the OpenACC programming model. 
The current public release\footnote{\url{https://gitlab.com/PLUTO-code/gPLUTO}} provides modules for magneto-hydrodynamics (MHD), and its relativistic extensions \citep[with the usage of Taub–Matthews equation of state;][]{Taub1948,Mignone2007_TAUB}, implemented in different geometries with Runge–Kutta time stepping. 
For the present study, the adoption of \texttt{gPLUTO} was essential as simulating jet propagation up to $\sim 5$~Mpc required computational domains containing upto 20 billion active cells, a scale at which GPU acceleration is indispensable for both runtime efficiency and memory throughput, while retaining the validated accuracy of the \textsc{PLUTO} framework.

\texttt{gPLUTO} solves the set of following relativistic-MHD equations, which in covariant form read
\begin{equation}
\begin{split} \label{eq:1}
    \partial_{\kappa} (\rho u^{\kappa}) = 0  \,,\\
    \partial_{\kappa} (w u^{\kappa} u^{\delta} - b^{\kappa} b^{\delta} +p g^{\kappa \delta}) = 0 \,, \\
    \partial_{\kappa} (u^{\kappa} b^{\delta} - u^{\delta}b^{\kappa}) = 0 \,,
\end{split}
\end{equation}
where $\kappa$, $\delta$ $= (0,\,1,\,2,\,3)$ \citep{Rossi2017}. In this formulation, $\rho$ denotes the rest-mass density and $u^{\kappa} = \Gamma(1,\mathbf{v})$ is the four-velocity, where $\Gamma$ is the Lorentz factor of the flow. 
The magnetic field four-vector $b^{\kappa}$ is defined in terms of the laboratory-frame magnetic field $\mathbf{B}$ as $(b^0, \mathbf{B}/\Gamma + b^0 \mathbf{v})$. 
The quantities $w$ and $p$ represent the total enthalpy and total pressure, respectively. 
Throughout this work, we assume a flat spacetime metric $g^{\kappa\delta} = \mathrm{diag}(-1,\,1,\,1,\,1)$.

The simulations were carried out using a $2^{\rm nd}$-order accurate spatial discretization based on linear reconstruction, coupled with the HLLC Riemann solver. 
The divergence-free condition of the magnetic field was enforced in the computational domain using the divergence-cleaning scheme of \citet{Dedner2002}. 
All simulations were executed on the Leonardo high-performance computing system operated by CINECA\footnote{\texttt{CINECA:} \url{https://www.hpc.cineca.it/systems/hardware/leonardo/}}. Computations were performed on the Booster partition, where each compute node is equipped with 4 GPUs and 32 CPU cores \citep{Tursini2023}. 
Depending on the specific simulation setup described below, we employed between 100 and 121 nodes, with typical wall-clock runtimes of approximately 17 hours per run. 
A recent study by \citet{Giri2026} presents a detailed performance benchmarking of the GPU-enabled PLUTO code for jet simulations, demonstrating substantial runtime speed-ups and validating its user readiness for such demanding simulations.

\subsection{Ambient medium configuration}\label{Sec:Ambient medium configuration}
%%%%%%%%%%%%%%%%%%%%%%%%%%%%%%%%%%%%%%%%%%%%%%%%%%%%%%%%%%%%
GRGs with megaparsec-scale extents are preferentially found to inhabit low-density regions of the cosmic web \citep[e.g., about 60\% in poor galaxy groups and about 10\% in filaments, sheets or voids;][]{Oei2024_7Mpc, Sankhyayan2024}. 
In particular, the most extreme GRGs show a marked preference for environments associated with galaxy filaments and poor groups \citep{Oei2022_5Mpc,Oei2024_7Mpc}. 
Such low-density surroundings play an important role in their growth, as the reduced ambient gas content imposes less resistance to jet propagation and lowers the likelihood of jet disruption through matter entrainment \citep{Rossi2008}.

\begin{figure*}
    \centering
    \includegraphics[width=0.95\textwidth]{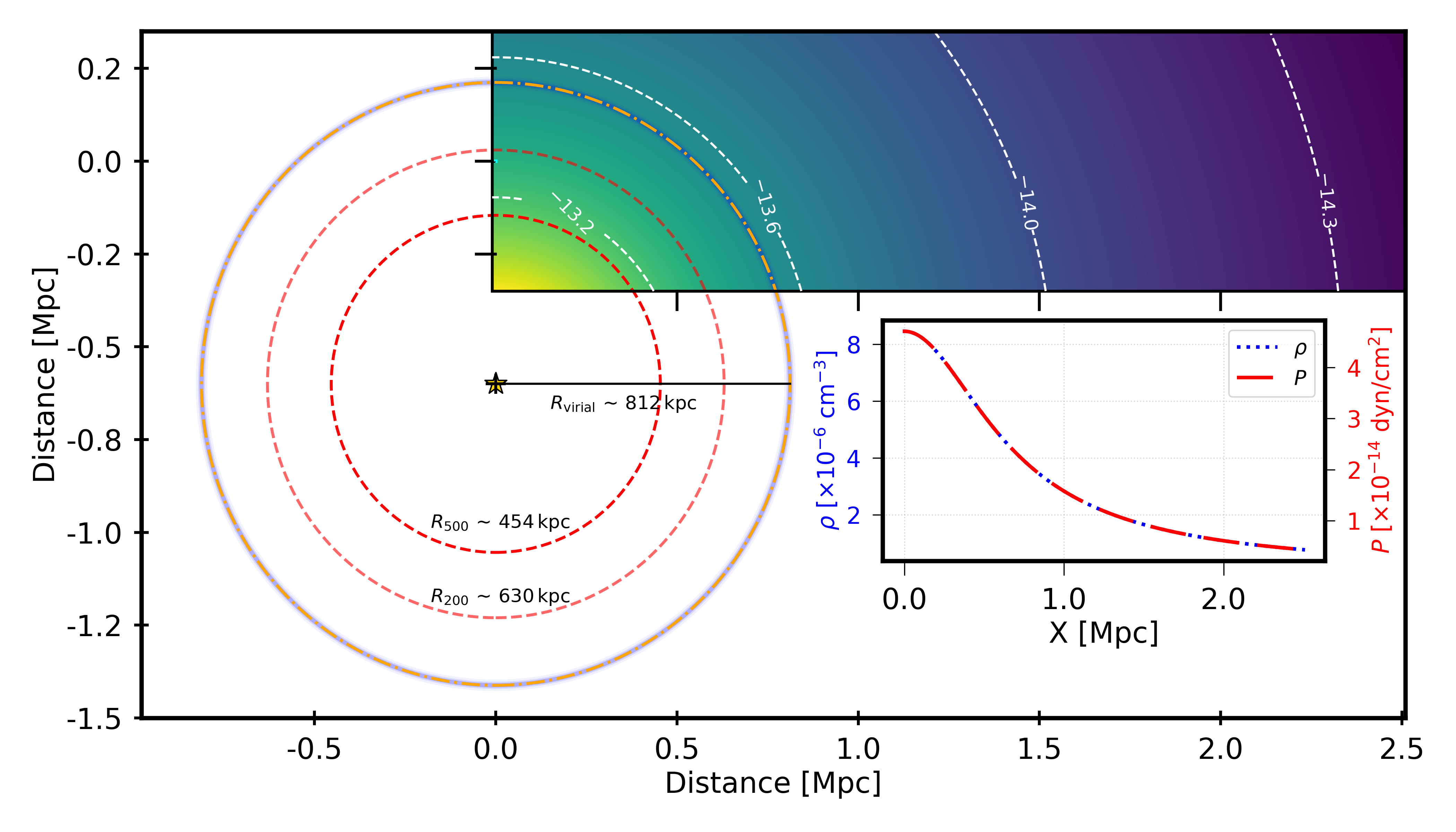}
    \caption{Initial setup of the simulation box (representative colourmap) for the GRG jet study. 
    The colormap shows the 2D density (in cm$^{-3}$) slice-cut distribution (log$_{10} \ \rho$ in the $x-y$ plane) with overlaid pressure contours (log$_{10} P$ in dyn cm$^{-2}$). The tiny cyan dot at the center ($0, 0$) represents the jet injection region. Circles indicate the group virial radius ($R_\mathrm{virial (100)}$, orange hued circle), $R_{200}$ (faded red dashed circle) and $R_{500}$ (red dashed circle). The star and plus symbols mark the center of the galaxy group. The inset shows 1D variations of density (blue dotted) and pressure (red dashed) along the jet injection (location: $x, y \equiv 0,0$; 600-kpc outskirt to the core) axis (along $x-$direction) to illustrate the ambient medium profile. The values presented here are chosen based on recent observational indications of GRG environments.}
    \label{Fig:setup}
\end{figure*}

Following \citet{Giri2025_GRGSim}, we modeled the ambient environment as an unmagnetized spherical galaxy group described by a King’s $\beta$-profile ($\rho(r)$) with a central gas density of $\rho_0 = 0.001\ \mathrm{amu \ cm^{-3}}$, core radius $r_c = 33$ kpc and $\beta = 0.55$,
\begin{equation}\label{Eq:Ambient_prof}
\rho(r) = \rho_0 \left[1 + \left(\frac{r-r_0}{r_c}\right)^2 \right]^{-\frac{3}{2}\beta}.
\end{equation}
These ambient parameters are motivated by the study of \citet{Ineson2015}, which established representative ranges based on a sample of 55 radio-loud AGN environments. An equilibrium intragroup medium temperature $T_g$ of $\sim 1.1\ \mathrm{keV}$ \citep{Sun2009,Lovisari2015} is obtained by assigning a pressure profile, $P(r)$, that follows the gas density distribution, $\rho(r)$ \citep[see Fig.~\ref{Fig:setup}; also,][] {Sun2011}. Hydrostatic equilibrium is enforced by incorporating a gravitational acceleration term in Eq.~(\ref{eq:1}), derived from the condition $\nabla P = \rho\,\mathbf{g}$.

For the adopted density profile, the integrated enclosed baryonic mass within \(2\ \mathrm{Mpc}\) is \(\sim 2.8 \times 10^{12}\ M_{\odot}\). For reference, this corresponds to a total mass of \(\sim 3.2 \times 10^{13}\ M_{\odot}\) when adopting a baryon fraction (baryonic-to-total mass ratio, including dark matter) of \(f_b = 0.09\), which we note represents the modal (binned mode) value of the baryon fraction distribution reported for group-scale systems in \citet{Sun2009}. We emphasize that this estimate provides only an order-of-magnitude comparison with standard group environments \citep[e.g.,][]{Vikhlinin2006,Humphrey2006,Paul2017} and systems hosting extended radio outflows and radio-loud jets \citep[e.g.,][]{Giacintucci2011,Nardini2013,Ineson2015}, rather than to match any specific observed cosmic system.

\begin{figure}
    \centering
   \includegraphics[width=\columnwidth]{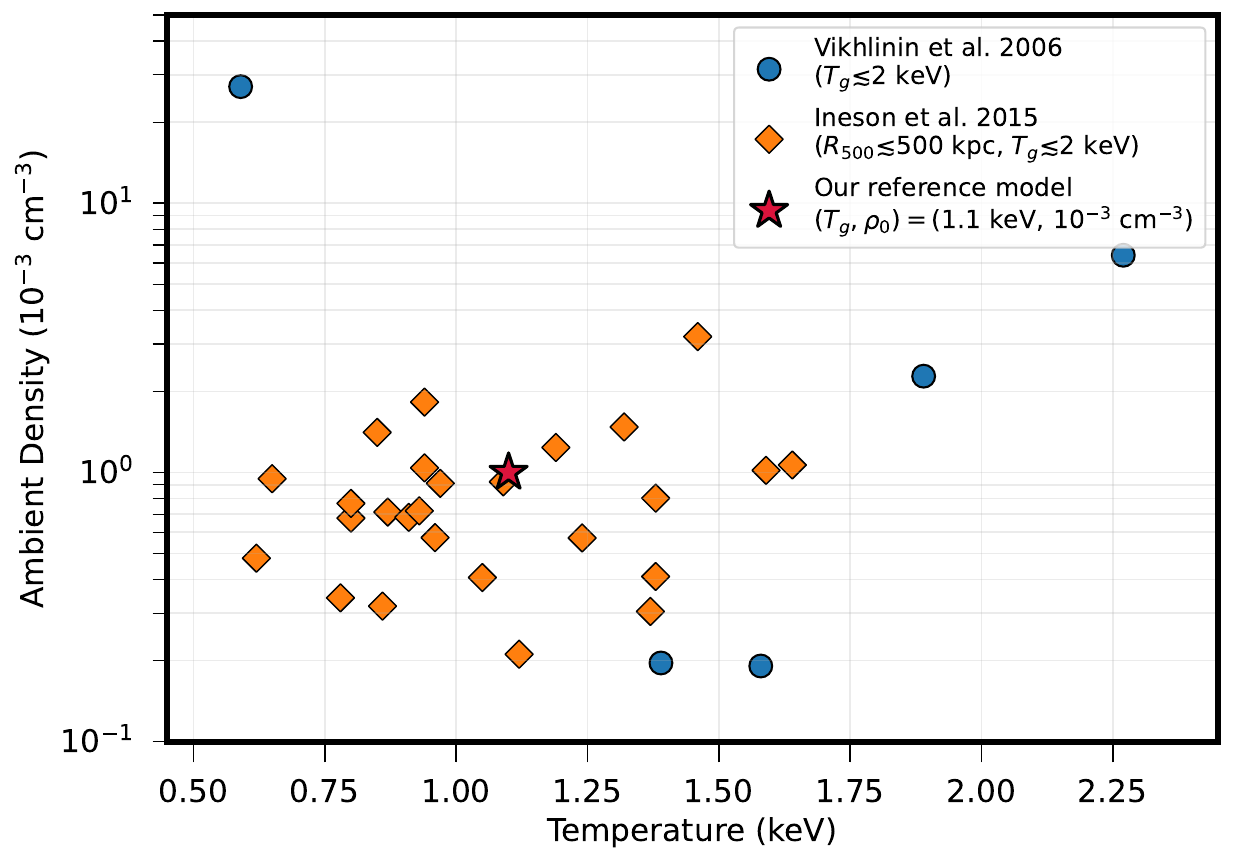}
    \caption{Central density ($\rho_0$) adopted in this study is compared with observationally inferred densities of group-scale environments, including relaxed galaxy groups and groups hosting radio jets, supporting the physical plausibility of our assumed ambient medium.}
    \label{Fig:Rho_0_comparison}
\end{figure}

The characteristic halo properties are derived self-consistently from the adopted gas density profile. In particular, we compute the overdensity radii \(R_{100}\) (virial radius), \(R_{200}\), and \(R_{500}\), defined as the radii within which the mean enclosed density equals \(100\), \(200\), and \(500\) times the critical density of the Universe at redshift \(z=0\), respectively. This yields \(R_{500} \sim 454\ \mathrm{kpc}\), \(R_{200} \sim 630\ \mathrm{kpc}\), and \(R_{100} \sim 812\ \mathrm{kpc}\). We further verify that the thermodynamic state of the modeled intragroup medium is consistent with observed scaling relations. Observational studies have shown that the mass–temperature (\(M_{500}\)–\(T_g\)) relation follows a robust scaling of the form \(M_{500} \propto T_g^{\alpha}\), with \(\alpha \sim 1.65\) across a range of group-scaled overdensity definitions \citep{Lovisari2015}. For our system with temperatures of order \(\sim 1.1\ \mathrm{keV}\), these relations predict halo masses of \(M_{500} \sim 2.6 \times 10^{13} M_{\odot}\), in good agreement with the values inferred in our setup. Furthermore, we present a visual comparison of the central density of our group-scale cosmic structure ($\rho_0$; Eq.~\ref{Eq:Ambient_prof}) with those measured in relaxed galaxy groups ($T_g \lesssim 2$ keV; \citealt{Vikhlinin2006}) and in galaxy groups hosting jetted radio sources with $R_{500} \lesssim 500$ kpc and $T_g \lesssim 2$ keV (\citealt{Ineson2015}). The close agreement between these systems and our modeled environment is showcased in Fig.~\ref{Fig:Rho_0_comparison}.

These consistencies provide independent validations that the adopted thermodynamical structure in our study represents a realistic group-scale environment.

At a higher redshift of $z=0.24674$ \citep[e.g., Alcyoneus][]{Oei2022_5Mpc}, the corresponding critical density of the Universe increases by a factor $\sim 1.25$, leading to a mild rescaling of the overdensity radii, with $R_{500} \simeq 420$ kpc, $R_{200} \simeq 581$ kpc, and $R_{100} \simeq 748$ kpc, consistent within $\sim10\%$ of the $z=0$ values used in our simulation setup.

In Eq.~(\ref{Eq:Ambient_prof}), the parameter $r_0$ accounts for a spatial offset, whereby the center of the galaxy group is displaced by 600 kpc along the negative $y$-axis in the Cartesian frame (Fig.~\ref{Fig:setup}). 
This setup allows the jet to be launched from the origin $(0,0,0)$ of the simulation domain \citep[see, `edge' case,][]{Giri2025_GRGSim}. 
This design ensures that the jet is injected from the outskirts of the galaxy group, thereby probing its propagation along the group’s periphery and subsequent evolution into a low-density \citep[$< 10^{-5}$ cm$^{-3}$;][]{Stuardi2020}, void-like environment representative of the large-scale cosmic surroundings of `extreme' GRGs \citep{Oei2022_5Mpc,Oei2024_7Mpc}. In Appendix~\ref{Sec:Influence of the host galaxy and the global intragroup medium}, we additionally present a complementary simulation in which the jet is injected from the centre of the galaxy group, in order to quantify the extent to which the group environment influences jet propagation and morphology.

\subsection{Jet injection configurations}
%%%%%%%%%%%%%%%%%%%%%%%%%%%%%%%%%%%%%%%%%%%%%%%%%%%%%%%%

We performed two jet configurations (as described below) to test how variations in thrust and collimation affect jet stability against both internal and external destabilizing effects for these large-scale GRGs.
While their kinetic powers differ by about an order of magnitude, both cases lie within the powerful FR II regime, allowing us to focus on the dynamical role of jet parameters rather than exploring fundamentally different energetic classes \citep{Fanaroff1974,Mingo2019}. 
The kinetic power of the injected jet is defined as
\begin{equation}
    Q_{\mathrm{j}} = \pi r_{\mathrm{j}}^{2} v_{\mathrm{j}} \, \Gamma \, (\Gamma - 1) \, \rho_{\mathrm{j}} c^{2},
    \label{Eq:JetPower}
\end{equation}
where $r_{\mathrm{j}}$ is the jet radius, $\Gamma = (1-v_{\rm j}^2)^{-1/2}$ is the bulk Lorentz factor of the flow, $v_{\mathrm{j}}$ is the jet velocity derived from $\Gamma$, $\rho_{\mathrm{j}}$ is the rest-mass density of the jet, and $c$ is the speed of light. 

In both simulations, the jets are injected from a small circular nozzle centered at ($0,0,0$) and directed along the positive $x-$axis (Fig.~\ref{Fig:setup}). 
Thereby, we distinguish between two configurations labeled as `\textit{Case~A}' and `\textit{Case~B}', which are discussed below and summarized in Table~\ref{Tab:jet_params}.  

\begin{table*}
\centering
\caption{Jet injection parameters adopted in the two simulation setups used in this study (keeping the ambient medium configuration unaltered).}
\label{Tab:jet_params}
\begin{tabular}{lcccccccc}
\hline\hline
Simulation label & $\Gamma$ & $r_{\mathrm{j}}$ [kpc] & $\rho_{\mathrm{j}}/\rho_0$ & $\sigma$ & $Q_{\mathrm{j}}$ [erg s$^{-1}$] & Description & Domain [Mpc$^3$] & Grid cube\\
\hline
`\textit{Case A}' & $7$  & $2.5$ & $10^{-5}$       & $0.01$ & $ 3.4 \times 10^{45}$ & Standard powerful FR-II jet & $2.52 \times 0.847 \times 0.847$ & $5040 \times 1694 \times 1694$\\
`\textit{Case B}' & $10$ & $2.0$ & $5 \times 10^{-5}$ & $0.10$ & $ 2.4 \times 10^{46}$ & High-thrust, collimated jet & $2.52 \times 0.70 \times 0.70$ & $6300 \times 1750 \times 1750$\\
\hline
\end{tabular}
\begin{tablenotes}
\small
\item \textbf{Notes.} $\Gamma$ is the bulk Lorentz factor, $r_{\mathrm{j}}$ is the jet injection radius, $\rho_{\mathrm{j}}/\rho_0$ is the jet-to-core density contrast (with $\rho_0 = 0.001$ amu cm$^{-3}$), $\sigma$ is the magnetization parameter, and $Q_{\mathrm{j}}$ is the kinetic jet power (Eq.~\ref{Eq:JetPower}). We model only one side of the jet, with the simulation domain (in Mpc$^3$) configured to optimally capture the jet–lobe structure while minimizing boundary effects; in all cases, the grid distribution ensures that the jet injection radius spans five cells.
\end{tablenotes}
\end{table*}

\textit{\textbf{\underline{Case~A}}} adopts a jet Lorentz factor of $\Gamma = 7$, motivated by our recent work \citep{Giri2026}, which demonstrates that such a choice of parameter is essential to produce jets on scales of $\approx 3$~Mpc. 
The jet is injected with a radius of $r_{\mathrm{j}} = 2.5$~kpc and with a density of $\rho_{\mathrm{j}} = 10^{-5}\rho_0$, yielding a kinetic power of $Q_{\mathrm{j}} \sim 3.4 \times 10^{45}$ erg s$^{-1}$. This configuration corresponds to a standard, yet powerful, FR~II-type radio jet \citep{Birzan2004}.  

\textit{\textbf{\underline{Case~B}}}, on the other hand, is designed to probe the robustness of jet propagation with higher thrust and enhanced collimation. 
For this purpose, the Lorentz factor is increased to $\Gamma = 10$ and jet density is $\rho_{\mathrm{j}} = 5 \times 10^{-5}\rho_0$, while the injection radius is reduced to $r_{\mathrm{j}} = 2.0$~kpc. 
This combination results in a higher kinetic power of $Q_{\mathrm{j}} \sim 2.4 \times 10^{46}$ erg s$^{-1}$.  

These two jet configurations are thus designed to explore the robustness of the jet parameter space in the `extreme' regimes of radio galaxies, testing how variations in Lorentz factor, injection radius, density contrast and resulting kinetic power \citep[in accordance with observational requirements;][]{Machalski2008,Oei2024_7Mpc} influence jets' growth on scales of $\sim 5$ Mpc.
We note here that the presence of highly relativistic jet spines with Lorentz factors up to $\Gamma \sim 10$ is not unreasonable \citep{Giovannini2001,Dubey2023} and has been cited for GRGs as well. 
For example, \citet{Laing2015} proposed a two-component structure consisting of a fast central spine ($\Gamma \sim 10$) embedded within a slower sheath while studying the prototypical GRG NGC~6251 with projected extent of 1.7 Mpc \citep{Cantwell2020}. 

In addition to bulk kinematics, the jet magnetic field configuration plays a crucial role in determining jet stability, although it makes a subdominant contribution to the (relativistic) jet power. 
In our simulations, we prescribe a toroidal magnetic field through the magnetization parameter $\sigma$, defined as the ratio of Poynting flux $B_j^2/4\pi$ to jet enthalpy flux $\Gamma\rho_j h_j$ \citep[similar to][]{Mukherjee2020,Rossi2017}. 
\begin{equation}
\sigma = \frac{B_j^2}{4\pi \Gamma^2 \rho_j h_j}, \qquad
B_y = B_j\, \mathcal{R} \sin\phi, \quad B_z = -B_j\, \mathcal{R} \cos\phi.
\end{equation}
where $B_y$ and $B_z$ are the toroidal magnetic-field components in our Cartesian 3D domain, perpendicular to the jet flow (jet flow fixed along the $x$-axis). Here, $\mathcal{R}$ and $\phi$ denote the polar coordinates in the $y$--$z$ plane, defined within the jet injection region (see next subsection). The magnetic field configuration is consistent with that adopted in earlier works \citep{Rossi2017, Giri2026_JetDir}. In our simulations, we adopt an initially purely toroidal magnetic field configuration, motivated by both physical and numerical considerations. Toroidal fields are commonly regarded as a sensible approximate configuration in relativistic jets, because the toroidal component is typically expected to dominate over the poloidal one \citep[e.g.,][]{Mizuno2015,Dubey2023}. This choice is also consistent with previous studies \citep[e.g.,][]{Mignone2010}, which show that relativistic jets with predominantly toroidal fields capture the essential features of jet dynamics at larger scales.

For `\textit{Case~A}', we adopt a modest value of $\sigma = 0.01$.
For `\textit{Case~B}', motivated by the higher Lorentz factor ($\Gamma = 10$), indicating increased thrust, we increase the magnetization to $\sigma = 0.1$. 
This latter choice reflects the expectation that powerful, highly relativistic jets are more resilient to instabilities \citep{Mukherjee2020}, allowing us to probe the role of magnetic fields in stabilizing the flow and mitigating matter entrainment \citep{Rossi2024}.

\subsection{Simulation box and boundaries}
%%%%%%%%%%%%%%%%%%%%%%%%%%%%%%%%%%%%%%%%%%%%%%%%%%%%%
Assuming symmetry between the jet and the counter-jet on a global size of $\sim 5$ Mpc radio structure, we restrict our modeling to a one-sided jet propagation for computational requirements.
Since the counter-jet is expected to follow a symmetric spatio-temporal evolution \citep[under the consideration that extreme jets are likely to be nearly in the plane of the sky;][]{Oei2023_lenghDist}, this approach significantly reduces the computational cost while still capturing the essential dynamics. 
Even with only half the jet pair simulated, the resulting setup represents one of the most computationally demanding jet simulations to date. 

The computational domain extends for $2.5$ Mpc along the jet axis (positive $x$-direction; Fig.~\ref{Fig:setup}). 
We resolve the jet radius with 5 grid cells (10 cells across the injection diameter), which should be enough to capture jet instabilities \citep{Mignone2013,Mukherjee2020}.
In `\textit{Case A}' (jet radius of 2.5 kpc), the grid comprises $5040 \times 1694 \times 1694$ cells, corresponding to a physical volume of $2.52 \times 0.847 \times 0.847$ Mpc$^3$. 
In `\textit{Case~B}' (jet radius of 2 kpc), the grid distribution is $6300 \times 1750 \times 1750$ cells, giving a volume of $2.52 \times 0.70 \times 0.70$ Mpc$^3$ (summarized in Table~\ref{Tab:jet_params}).
To the extent of our knowledge, this is the largest relativistic jets simulation performed so far.

The boundary conditions are taken to be reflective outside the jet nozzle at the lower $x$-boundary plane, mimicking the plasma contribution from the unmodeled counter-jet, while all other faces adopt outflow conditions. 
The jet injection zone is defined as a cylinder of radius $r_j$ (2.0 or 2.5 kpc, depending on the case) and length $r_j$, centered at $(0,0,0)$. 
All simulations are carried out in dimensionless units, converted back to physical values using $L_0 \equiv 100$ kpc (length), $\rho_0 \equiv 10^{-3}$ amu cm$^{-3}$ (density), and $v_0 \equiv c$ (velocity as light speed), consistent with \citet{Giri2025_GRGSim}. 
In addition, a passive scalar tracer ($\mathcal{T}_j$) is injected and advected with the jet flow. 
Its values range from 1 (indicating a grid filled with jet plasma) to 0 (indicating a grid filled with ambient medium), enabling quantitative tracking of mixing and jet–ambient separation \citep[e.g.][]{Mignone2013}.

%%%%%%%%%%%%%%%%%%%%%%%%%%%%%%%%%%%%%%%%%%%%%%%%%%%%%%%%%%%%%%%%%%%
\section{Morphological and Temporal Evolution}\label{Sec:Morphological and Temporal Evolution}
%%%%%%%%%%%%%%%%%%%%%%%%%%%%%%%%%%%%%%%%%%%%%%%%%%%%%%%%%%%%%%%%%%%

\begin{figure*}
    \centering
    \includegraphics[width=0.95\textwidth]{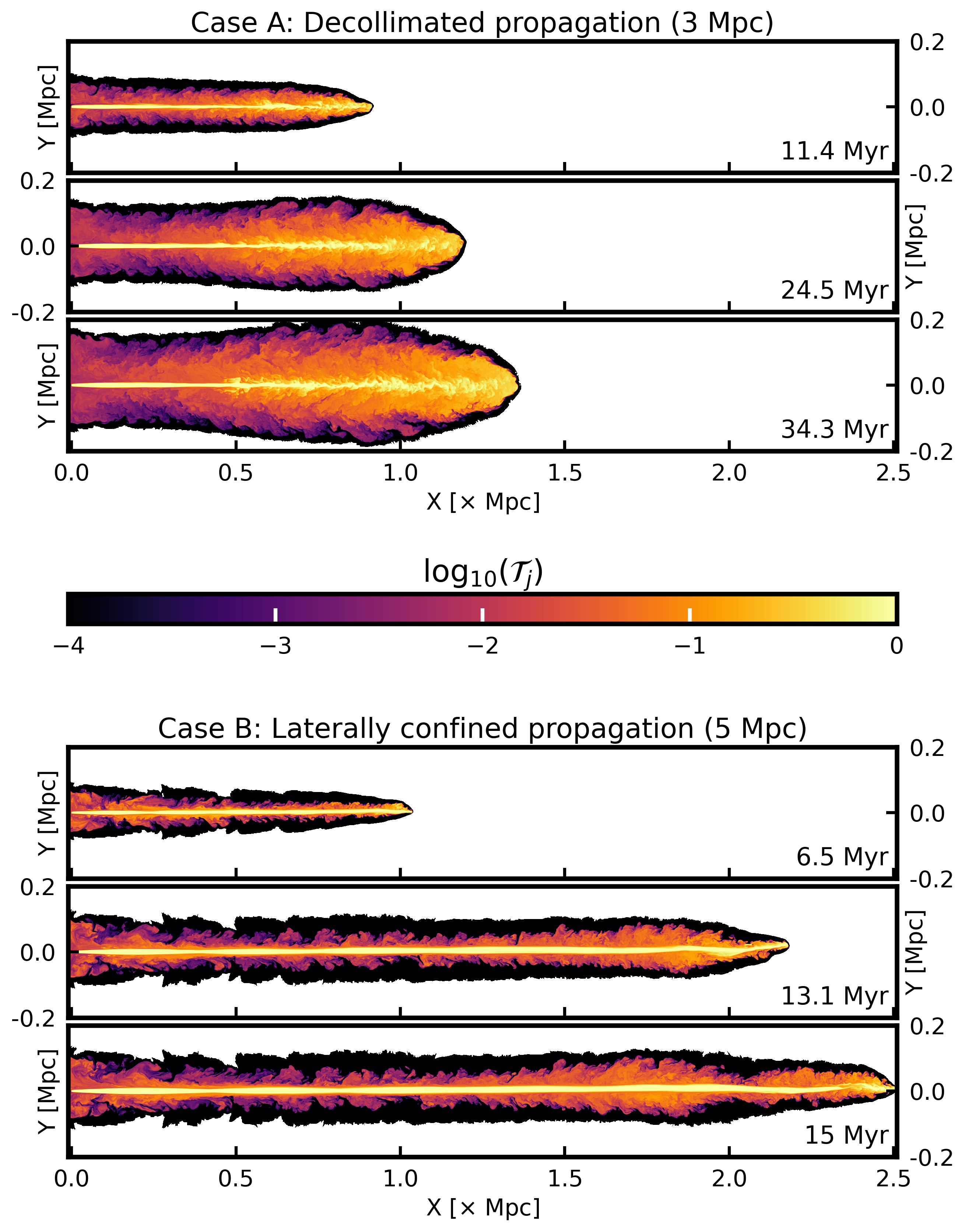}
    \caption{Two-dimensional ($x-y$, $z=0$) jet tracer distributions for `\textit{Case A}' (\textit{top}) and `\textit{Case B}' (\textit{bottom}), illustrating the spatio-temporal evolution of jets producing GRGs. The results highlight that both jet collimation and thrust are essential for enabling GRGs to reach extreme scales. The top rows show the benchmark case in which jet decollimation leads to enhanced dissipation and limits the bi-directional extent to $\sim 3$ Mpc. In contrast, the bottom rows illustrate a more tightly collimated jet that rapidly propagates to $\sim 5$ Mpc (bi-directional extent) in less than half the evolutionary time than `\textit{Case A}', before developing mild wiggling and undulations, while still maintaining rapid propagation.}
    \label{Fig:Case_evolution}
\end{figure*}

The temporal evolution of our two simulations (Table~\ref{Tab:jet_params}) exploring the regime of `extreme' jet propagation is shown in Fig.~\ref{Fig:Case_evolution} using the tracer (passive scalar quantity: $\mathcal{T}_j$).

`\textit{Case A}' (standard powerful jet), demonstrates rapid propagation to $\sim1$ Mpc within 11.4 Myr, maintaining a narrow, collimated jet–cocoon structure. 
However, once the jet begins to decollimate (at $\sim 11.4$ Myr), the jet flow transitions from a cylindrical to a more diffuse morphology. 
Despite continued thrust at the jet head, this decollimation in the jet flow significantly slows the advancement along the propagation axis and enhances lateral expansion, leading to the formation of a lobe-like GRG with a maximum extent along the propagation direction of $\sim 1.4$ Mpc (around 35 Myr), consistent with the growing population of $\sim 3$ Mpc-sized radio galaxies.

`\textit{Case B}' (more powerful jet) maintains its collimation for considerable longer distance, reaching $\sim2.2$ Mpc in just $13.1$ Myr \citep[see relevant spectral ageing studies of GRGs;][]{Jamrozy2008,Sebastian2018}. 
Beyond this stage, the jet begins to exhibit mild oscillations that signal the onset of slow decollimation and likely instabilities (discussed further in \S\ref{Sec:Dynamics of Jet Propagation}).
Nevertheless, the jet thrust remains efficiently delivered to the head, allowing the system to extend to $\sim2.5$ Mpc in $15$ Myr—demonstrating that `extreme' GRGs can be explained within the standard jet paradigm. 
Such ages, of order $\sim10^7$ yr, contrast sharply with the $\gtrsim$ Gyr timescales previously inferred observationally for such sources \citep[e.g.,][]{Oei2022_5Mpc,Oei2024_7Mpc}, which would require sustained accretion onto the host AGN over cosmologically significant timescales. 
The persistence of collimation and directionality over those timescales is also non-trivial \citep{Nernonov2025}: even modest angular flickering of only a few degrees \citep[e.g., $\sim 3^{\circ}$, median value in restarting GRGs;][]{Dabhade2025}, would correspond to large ($\sim$350 kpc for `Porphyron') transverse displacements at these multi-Mpc extents, potentially driving decollimation in the jet propagation \citep[e.g.,][]{Guan2014,Cantwell2020,Dabhade2022}. 
Our results therefore suggest that the extreme sizes of GRGs can arise more naturally from efficient, collimated jet dynamics than from prolonged ($\gtrsim$ Gyr) lifetimes, consistent with other evolutionary estimates \citep[e.g.,][]{Machalski2008,Giri2025_GRGSim}.

\begin{figure}
    \centering
   \includegraphics[width=\columnwidth]{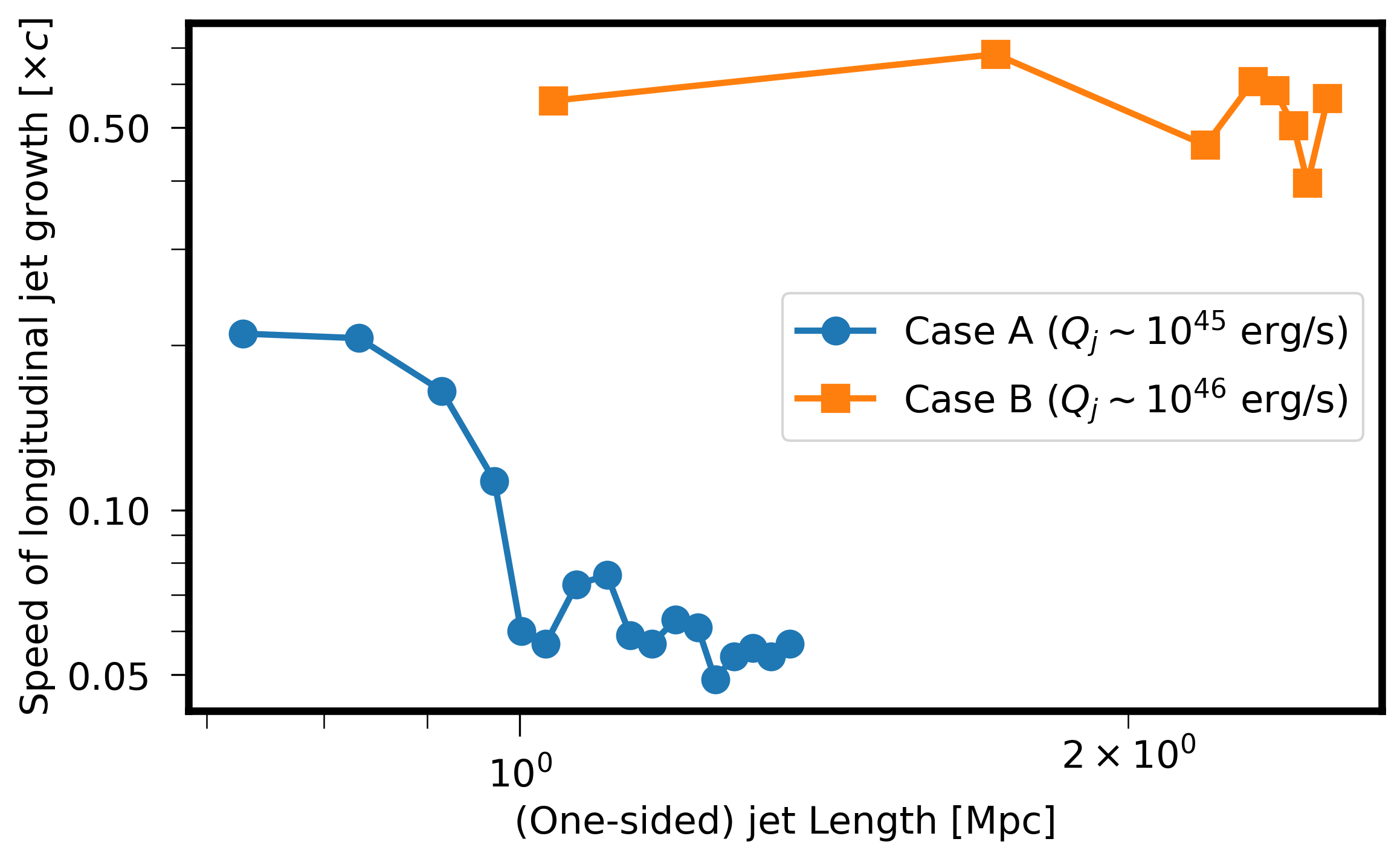}
    \caption{The axial growth speed of the jet structure (defined as the temporal increase in its total longitudinal extent; in units of the speed of light, $c$) as a function of the growing jet length (in Mpc) for the two simulation sets. The plot illustrates how the emergence of two different classes of GRGs is primarily governed by jet collimation and thrust, both of which are jointly encapsulated by the jet power.}
    \label{Fig:V-L}
\end{figure}

To further illustrate the effects of decollimation, Fig.~\ref{Fig:V-L} shows the evolution of longitudinal growth speed of the jet structure as a function of its growing length (mimicking the temporal evolution as well) for `\textit{Cases A}' and `\textit{B}'. 
Within our adopted jet power regimes, the two cases display distinct evolutionary behaviors. 
In `\textit{Case A}', the jet rapidly reaches $\sim 1$ Mpc with a lobe expansion speed exceeding $0.2 c$, producing a thin cocoon with an axial ratio of 9.2 (linear-to-lateral length ratio). 
Beyond this scale, the expansion slows significantly to $\sim 0.05c$, and the axial ratio decreases to 4.3, reflecting substantial lateral dissipation. 
In contrast, `\textit{Case B}', with higher thrust and better collimation, maintains a lobe expansion speed around $0.5c$, and its axial ratio increases from $9.5$ to $12.5$, indicating continued longitudinal growth and thinner lobes. 
The jet head speed shows only mild variations due to minor wiggling (Fig.~\ref{Fig:Case_evolution}), demonstrating that such jets possess the mechanical thrust neeeded to propagate to bilateral extents beyond 5 Mpc, offering insight into the formation of even larger scaled GRGs (Giri et al. (\textit{in prep.})). 
The axial ratio values reported above provide a quantitative basis for distinguishing between lobed and non-lobed GRGs, consistent with an observational distinction reported in \citet{Subrahmanyan1996} and \citet{Machalski2008}.

To place our results in an observational context, we compare the simulated longitudinal propagation speeds with reported advance speeds of radio galaxies across different size scales. Classical radio galaxies (median extent $\sim 100$ kpc) exhibit head advance speeds in the range $\sim 0.03c$--$0.2c$, with a median value of $\sim 0.1c$ \citep{Liu1992}. Giant radio galaxies show a comparable range, $\sim 0.03c$--$0.25c$, with a median $\sim 0.09c$ \citep{Jamrozy2008}. Even among the most extreme $\sim 5$ Mpc systems, such as J1420$-$0545 and 3C\,236, inferred advance speeds span $\sim 0.065c$--$0.163c$ \citep{Machalski2008}. In this context, the $\sim 0.5c$ longitudinal growth speed obtained for `\textit{Case B}' lies toward the upper envelope of observationally inferred values and exceeds the typical medians. However, it is not drastically disconnected from the highest reported speeds, particularly considering the substantial uncertainties inherent in spectral-ageing-based estimates. Furthermore, compact symmetric objects (CSOs), which provide some of the most direct hotspot proper-motion measurements, exhibit a broad range of expansion speeds from $\sim 0.04c$ up to $\sim 0.45c$ \citep{Polatidis2003,An2012}, demonstrating that such high advance speeds are not unprecedented in radio-loud AGN, albeit on much smaller spatial scales. Our future planned works will extend current numerical framework by exploring jet propagation across a broader range of environmental densities and large-scale cosmic web locations, enabling a more systematic assessment of how ambient conditions regulate the longitudinal growth rate of `extreme' GRGs.

Given that the jet--lobe structures are still evolving, it is instructive to compare the cocoon pressure against the surrounding environment, at their evolved stages. 
For the $\sim 3\,\mathrm{Mpc}$ GRG, we estimate a lobe pressure of $8.6 \times 10^{-13}\,\mathrm{dyn\,cm^{-2}}$, while for the $5\,\mathrm{Mpc}$ GRG the pressure is higher at $2.0 \times 10^{-12}\,\mathrm{dyn\,cm^{-2}}$. 
The latter is thus about $2.3$ times over-pressurized, consistent with its more elongated, non-lobed morphology. 
Relative to the ambient medium, the $3\,\mathrm{Mpc}$ and $5\,\mathrm{Mpc}$ jet structures remain over-pressurized, respectively, by a factor of $\sim 4.1$ and $\sim 3.6$. The values remain broadly comparable, considering that a larger fraction of pressure-driven energy is transferred to the ambient medium from the lobe in `\textit{Case B}' than in `\textit{Case A}' (further discussed in \S~\ref{Sec:Jet -- environment interconnection}).
At earlier times, this contrast was significantly higher: for the \textit{Case A} jet at $11.4\,\mathrm{Myr}$ the cocoon was $16.3$ times overpressured, while for the \textit{Case B} jet at $6.5\,\mathrm{Myr}$ the contrast was $14.4$. 
These results highlight two points: (a) cocoon pressure decreases as the system evolves to larger scales, suggesting that extreme GRGs may eventually reach pressure equilibrium with their environments, providing a potential probe of poorly explored large-scale baryonic structures such as the WHIM and cosmic voids \citep{Subrahmanyan2008,Malarecki2013,Oei2023_spiralHost}; and (b) in the present stage of our simulations, the lobes are still modestly overpressured compared to their environments.

\begin{figure*}
    \centering
    \includegraphics[width=0.95\textwidth]{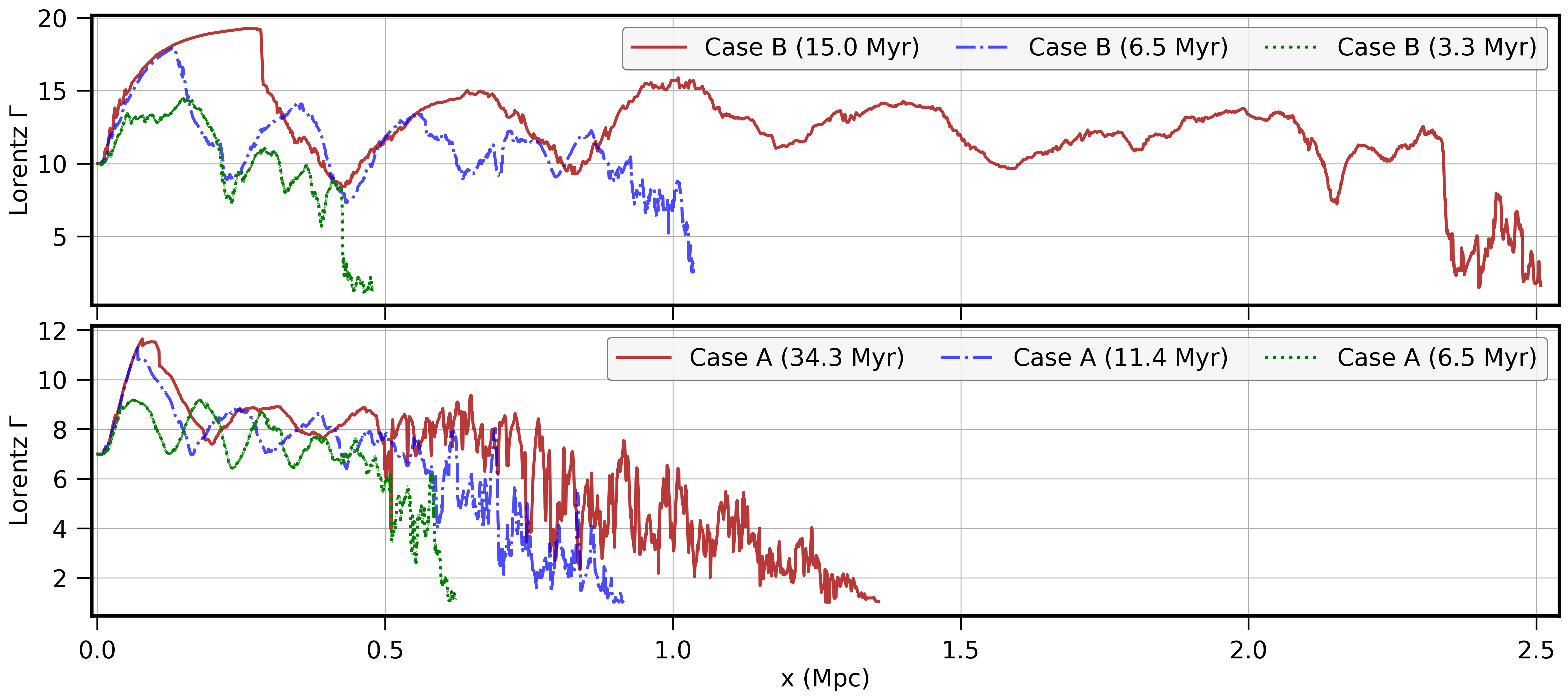}
    \caption{Longitudinal profiles of the Lorentz factor ($\Gamma$) along the jet spine for `\textit{Case A}' and `\textit{Case B}' at different evolutionary times. The contrast between the steep versus gradual dissipation near the jet head, together with the pronounced undulations of $\Gamma$ along the spine, highlights the different degrees of collimation, shock structure, and interaction with the surrounding cocoon in the two simulations.}
    \label{Fig:LorG_timeEvol}
\end{figure*}

A useful diagnostic of the jet spine, identified with jet tracer distribution $\mathcal{T}_j \geq 0.5$, is the evolution of the Lorentz factor $\Gamma$ along the axis as the system develops. Fig.~\ref{Fig:LorG_timeEvol} shows $\Gamma$ for different evolutionary stages in both simulations. 
In `\textit{Case B}', the jet behaves like a powerful, well-collimated flow, mirroring the behavior observed in several collimated GRG jets \citep[e.g.,][]{Saripalli1994,Sebastian2018,Cantwell2020}. 
Shortly after injection, the jet spine accelerates from $\Gamma \approx 10$ to nearly $\Gamma \approx 20$ as the jet over-expands into a comparatively under-pressured environment. 
This over-expansion triggers a sequence of recollimation shocks that impose a quasi-periodic modulation on the spine \citep{Mizuno2015,Costa2025}. 
Importantly, the amplitude of these oscillations increases with temporal evolution: as the cocoon is progressively populated by back-flowing jet material \citep[e.g.][]{Cielo2017}, plus its internal pressure drops due to expansion, thus the jet column becomes increasingly over-pressured, leading to stronger oblique shocks and more efficient rarefaction-driven acceleration between them. Despite the growing interaction region near the head, where $\Gamma$ drops sharply and strong backflows form, the jet column remains fast and only weakly affected by entrainment of ambient matter. 
The persistently steep $\Gamma$ gradient at the head therefore indicates largely non-dissipative, ballistic-like transport that could in principle propagate to even larger distances.

By contrast, `\textit{Case A}' shows a more gradual decline of $\Gamma$ along the column, signaling progressive loss of collimation and greater coupling with the cocoon and the ambient environment. 
At early times ($\approx 6.5$ Myr), the $\Gamma$ profile still resembles `\textit{Case B}', suggesting that the jet initially behaves as a very well focused outflow. 
However, as the system evolves ($\approx 11$ Myr), the oscillations dampen and the $\Gamma$ profile flattens, consistent with (ambient) mass loading and mixing that increase the jet's inertia and suppress strong recollimation shocks \citep[see,][for analogous GRGs]{Subrahmanyan1996}. 
The jet slows more uniformly, and the head advances less efficiently, in agreement with the reduced lobe expansion seen in Fig.~\ref{Fig:V-L}.
Even so, a weaker residual jet column persists near the base, with localized $\Gamma$ enhancements occasionally exceeding the injection value ($\Gamma \in [7,\,12]$
), allowing portions of the flow to travel significant distances before dissipating.

\begin{figure*}
    \centering
    \includegraphics[width=\textwidth]{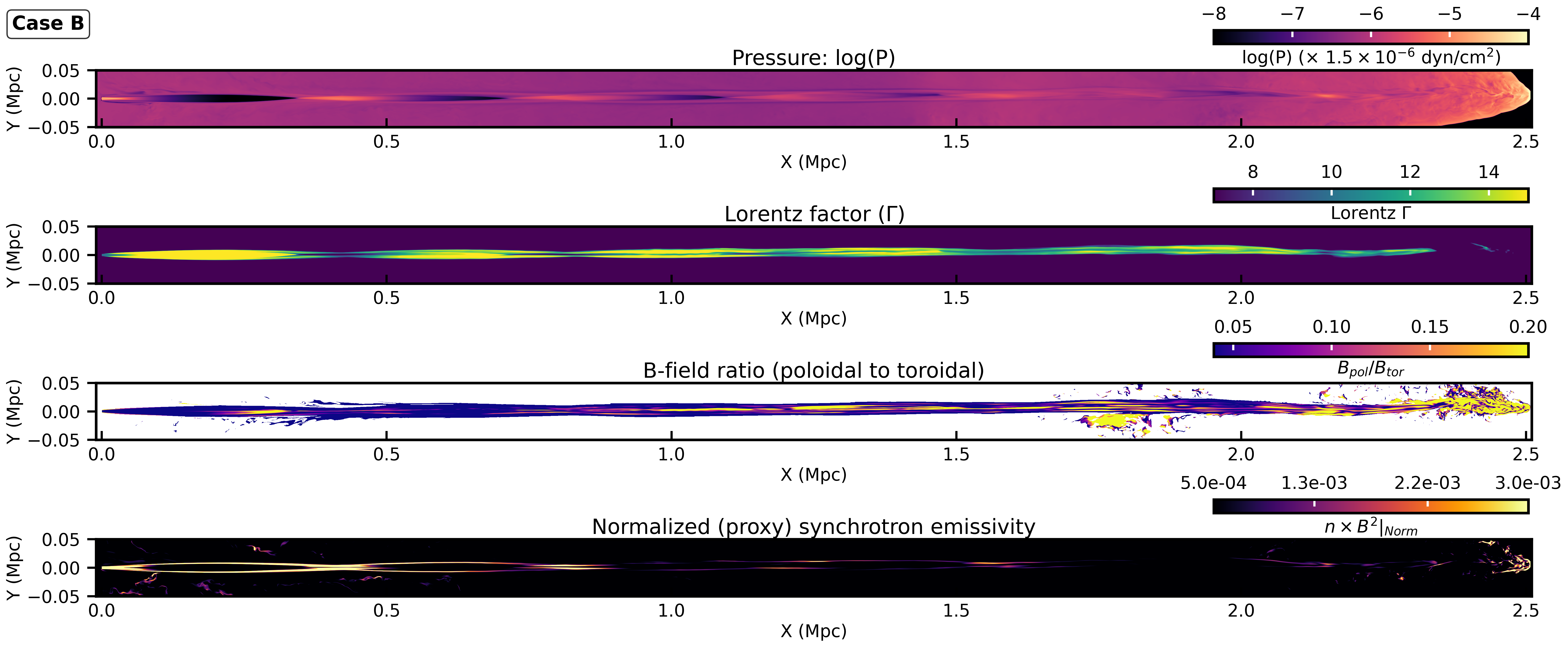}
    \caption{{\bf A.} Jet-spine diagnostic maps for `\textit{Case B}' at 15 Myr ($x,y$ slice, $z = 0$). Pressure, Lorentz $\Gamma$, magnetic-field geometry, and proxy emissivity illustrate a cocoon-confined, self-regulated jet, with shock-driven recollimation, poloidal-field amplification, and bright emission concentration at the base and head. The proxy emissivity ($\propto nB^2$), defined as the particle density multiplied by the magnetic energy density, is normalized and visualized using a restricted color scale chosen to optimally highlight the spatially extended emission.}
    \label{Fig:Jet_spine_B}
\end{figure*}

A clearer view of the above discussion can be obtained by examining 2D maps of pressure, Lorentz factor, magnetic-field geometry, and proxy synchrotron emissivity.
We first focus on `\textit{Case B}', since its spine remains comparatively well collimated (see, Fig.~\ref{Fig:Jet_spine_B}.A.). 
The pressure and Lorentz-factor panels reveal the expected anti-correlation: regions of enhanced pressure correspond to local deceleration, while rarefaction zones coincide with $\Gamma$ increases. 
This pattern traces a sequence of recollimation and rarefaction waves driven by the initial over-pressured injection relative to the environment (Fig.~\ref{Fig:Jet_spine_B}.A.). 
The resulting shocks redistribute energy between thermal and kinetic components, progressively smoothing out and becoming less pronounced as the jet equilibrates with the cocoon. 
Near the head, $\Gamma$ drops sharply and becomes asymmetric, reflecting strong interaction with the environment and the onset of turbulent backflows that inflate a pressure-confined cocoon around the beam. 
The cocoon therefore provides a quasi-uniform pressure bath that self-regulates the jet spine. 
The magnetic-field ratio map (Fig.~\ref{Fig:Jet_spine_B}.A.) shows that the jet is initially dominated by the toroidal component ($B_{\rm tor}$: following our injection condition) which becomes progressively stretched into the flow direction at recollimation sites, being converted into poloidal field ($B_{\rm pol}$). 
These poloidal patches appear primarily where shock compression and shear act on the beam.
This is consistent with the expectations for powerful jets in which the jet plasma and B-field effectively remain locked in, while the forward motion of the jet stretches the field along the propagation direction \citep{Laing1981}. 
Notably, where $B_{\rm pol}$ strengthens, the jet column becomes less wiggly, indicating a magnetically mediated damping of transverse fluctuations \citep{Biskamp1998}. This assumption breaks down near the jet head, where the collimated beam structure is disrupted and a poloidal component arises from lateral mass displacement associated with gradual decollimation.
Finally, the synchrotron proxy emissivity, assumed to be $\propto n B^2$ where $n$ is the (jet-fluid) particle density (in amu $\cdot$ cm$^{-3}$) and $B$ the field magnitude, is shown in the bottom panel of Fig.~\ref{Fig:Jet_spine_B}.A.. 
Bright emission is observed near the base produced by the strong oblique shocks, followed by a comparatively faint segment along the beam where the developing $B_{\rm pol}$ helps to streamline the flow, and a renewed brightening at the jet head where the interaction with the ambient medium is strongest. 
This pattern closely resembles the emission structure seen in the most extended radio galaxies \citep{Machalski2008,Oei2024_7Mpc}.

\begin{figure*}
    \ContinuedFloat
    \centering
    \includegraphics[width=\textwidth]{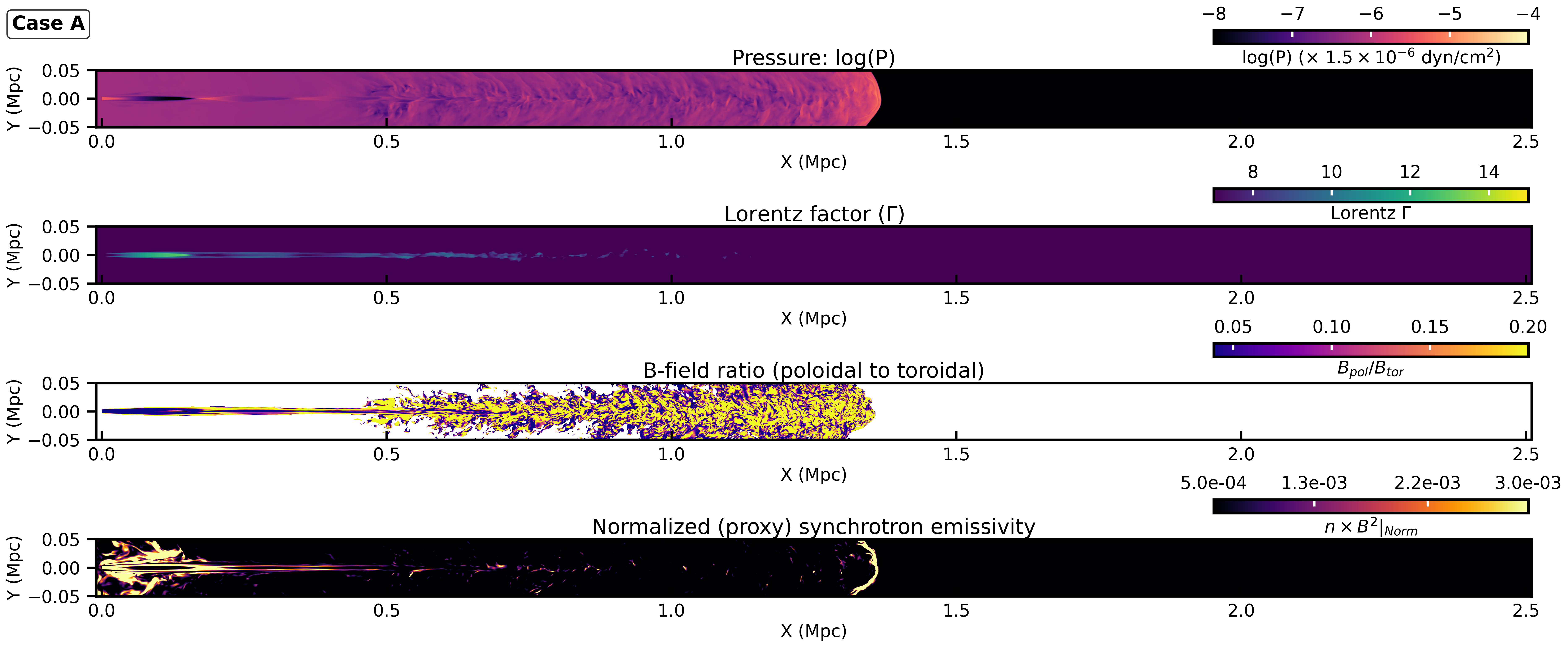}
    \caption{{\bf B.} \textit{Continued.} Jet-spine diagnostic maps for `\textit{Case A}' at 34.3 Myr ($x$–$y$ slice, $z = 0$). Pressure, Lorentz $\Gamma$, magnetic-field geometry, and proxy emissivity illustrate how a high-$\Gamma$ spine progressively loses collimation, dissipates into the cocoon, and forms a widespread terminal emission region. Proxy emissivity ($\propto nB^2$) is normalized and visualized using a restricted color scale chosen to optimally highlight the spatially extended emission.}
    \label{Fig:Jet_spine_A}
\end{figure*}

In contrast to `\textit{Case B}', the jet in `\textit{Case A}' dissipates more rapidly, but several notable features remain (see, Fig.~\ref{Fig:Jet_spine_A}.B). 
Near the jet base (up to $\sim 0.5$ Mpc), a series of prominent oblique shocks produce localized pressure enhancements and Lorentz factor variations, indicating an initially self-modulating jet spine embedded in a quasi-uniform cocoon. 
However, not far downstream, the spine begins to wiggle asymmetrically under the influence of turbulent structures in the cocoon, progressively perturbing the flow.
Although the jet fails to preserve a stable spine over the extreme GRG-like scales of interest, the early, well-collimated phase — characterized by relatively high $\Gamma$ (reaching $\sim 12$)— still allows the jet to propagate a considerable distance before breaking apart. 
The poloidal magnetic component grows downstream of the compression regions, similar to Case B, but its amplitude remains significantly lower, reflecting the weaker injected magnetic field at the jet base (lower by roughly an order of magnitude; Table~\ref{Tab:jet_params}; discussed in the next paragraph). 
Combined with the lower propagation speed of this case and the persistent cocoon turbulence \citep{Mukherjee2020}, the jet gradually dissipates, transitioning from axial transport to enhanced lateral expansion. This lateral dissipation drives the development of poloidal magnetic components primarily through advection.
In the mock emissivity map, these dynamics manifest as bright emission at the base (due to oblique shocks), a nearly vanishing, irregularly wiggling jet spine, and a terminal bright region where the jet interacts strongly with the ambient medium \citep[see, e.g.,][]{Oei2022_5Mpc}.

\begin{figure*}
    \centering
    \includegraphics[width=0.95\textwidth]{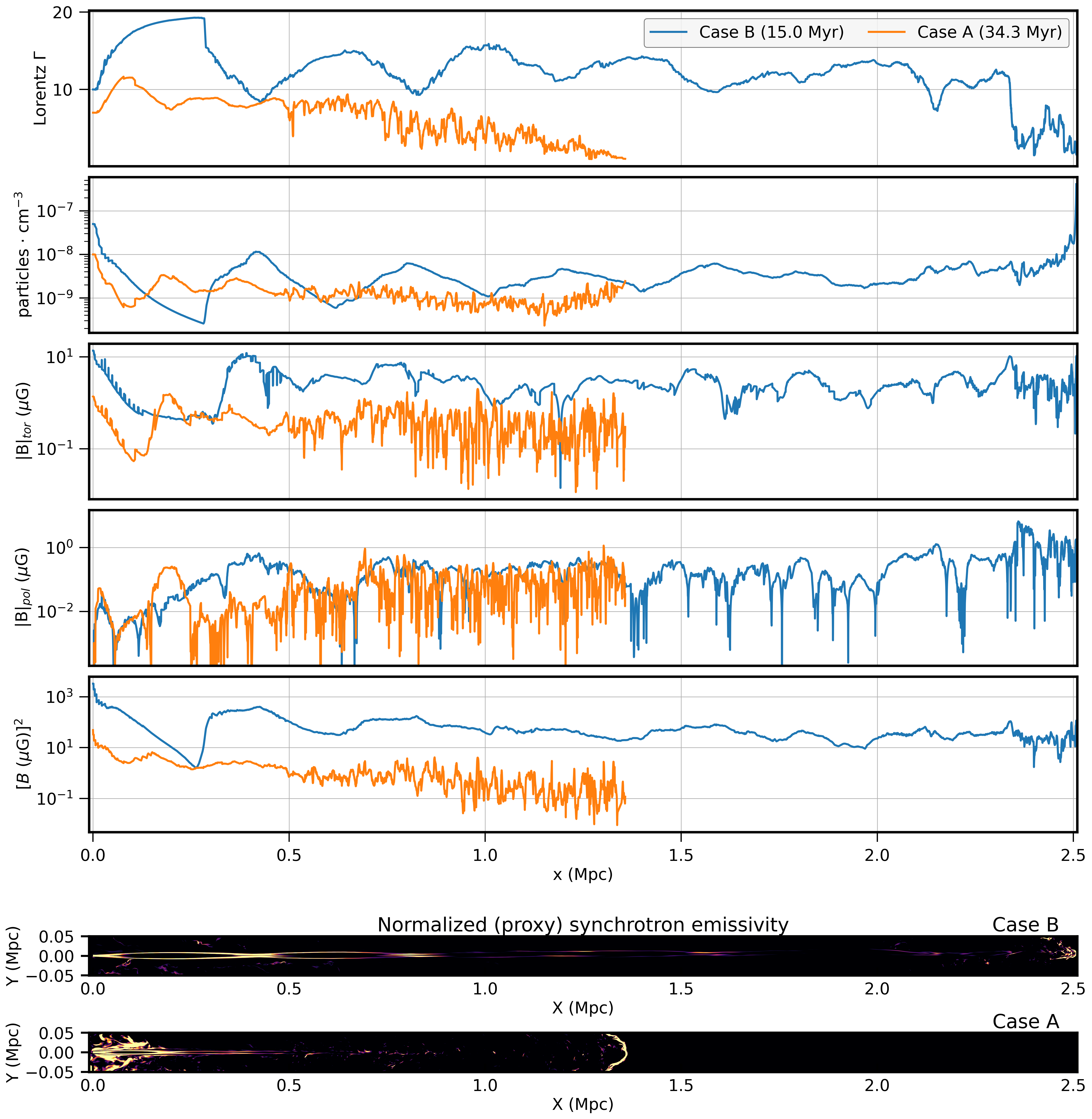}
    \caption{Comparison of `\textit{Case A}' and `\textit{Case B}' jet-spine profiles in 1D, showing how particle density and the toroidal and poloidal magnetic-field magnitudes vary along the spine, and how they correlate with the Lorentz factor. The profiles also highlight the enhanced emission near the jet base arising from simultaneous amplification of particle density and magnetic energy ($\propto nB^2$), modulated by the first few recollimation shocks. Downstream, the flattening of density and the comparatively plateau-like magnetic field variations lead to reduced emissivity along the spine, while strong interaction at the jet head produces a localized re-brightening. These profiles help identify the dynamics operating in the quasi-periodic oscillatory regions and further downstream, including compression/rarefaction behavior and progressive jet–cocoon interaction, illustrating how synchrotron emission is concentrated in `active' jet regions.}
    \label{Fig:1D_variations_AB}
\end{figure*}

To further refine the interpretation developed above in Figs.~\ref{Fig:Jet_spine_B}.A and  \ref{Fig:Jet_spine_A}.B, we examine the 1D profiles of several jet parameters in Fig.~\ref{Fig:1D_variations_AB}. 
For reference, the Lorentz factor is shown again, together with the particle density and the absolute values of the toroidal and poloidal magnetic-field components along the jet column. 
Unlike the ratio maps, these profiles reveal how the individual field components evolve in magnitude, allowing us to connect the dynamics more directly to the mock emission maps (providing a first-order description that does not incorporate more complex processes such as particle cooling and reacceleration). 
The compression and rarefaction sites identified in the $\Gamma$ profile are clearly mirrored in the particle density. 
These oscillations are more pronounced in `\textit{Case B}', consistent with its higher injected jet density and larger Lorentz factors (Table~\ref{Tab:jet_params}). 
Toward the jet head, both simulations exhibit a density increase and a sharp drop in $\Gamma$, signaling progressive mass loading from the ambient medium and a transition from ballistic propagation to a cocoon-regulated flow regime \citep{Leahy1984,Bromberg2011,Cielo2017}. 
This same region coincides with enhanced emissivity in the maps, reflecting the joint amplification of density and magnetic field. 
The toroidal component begins roughly an order of magnitude stronger in `\textit{Case B}' than in `\textit{Case A}'. 
It responds only weakly to the early compression–expansion cycles and soon settles into a relatively smooth, slowly varying plateau profile. This behavior indicates that the toroidal field is largely being advected with the flow rather than reorganizing the jet structure. 
In contrast, the poloidal component starts very weak but grows steadily downstream (starting near the jet injection region at values $\lesssim 10^{-2}$ and increasing to $\sim 10^{0}$ near the jet head), especially in `\textit{Case B}'. 
Because the vertical plot axis is logarithmic (Fig.~\ref{Fig:1D_variations_AB}), the apparently modest increase corresponds to a substantial amplification, spanning more than an order of magnitude. 
The nature of this growth indicates that it is produced by continuous shear and longitudinal stretching. 
Where $B_{\rm pol}$ strengthens, the jet column becomes less oscillatory, supporting the interpretation that the progressive development of poloidal field contributes to damping transverse perturbations and stabilizing the jet over large distances. As noted earlier, this effect is prominent within the well-collimated jet column, but not in the advecting, decollimating region where the poloidal field arises from lateral advection and no longer contributes to column support.

To make this connection explicit, we extend Fig.~\ref{Fig:1D_variations_AB}  to include the magnetic energy density ($\propto B^2$), together with the corresponding normalized proxy synchrotron emissivity maps (from Fig.~\ref{Fig:Jet_spine_B}.A and \ref{Fig:Jet_spine_A}.B). 
These additions clarify that the enhanced emission near the jet base arises from the simultaneous amplification of particle density and magnetic energy density, as captured by the elevated $nB^2$ values in this region. 
The emissivity near the jet base is locally enhanced by the first few recollimation shocks, which modulate both particle density and magnetic energy density, producing a sequence of brightness peaks. 
This indicates that basal emission is dynamically sustained by repeated compression rather than solely by high injection of $n$ and $B$ (Table~\ref{Tab:jet_params}). 
Downstream of the initial collimation zone, both simulations exhibit a significant flattening of the density profile and comparatively plateau-like magnetic field variations, leading to a sharp decline in emissivity and rendering the jet spine largely radio faint over extended distances. 
In contrast, at the jet head, strong interaction with the ambient medium increases especially density, leading to a localized re-brightening. 
Overall, the observable synchrotron emission is concentrated in regions of active compression and dissipation within the jet, rather than along its entirety, and also at the head location.

\section{Dynamics of Jet Propagation}\label{Sec:Dynamics of Jet Propagation}
%%%%%%%%%%%%%%%%%%%%%%%%%%%%%%%%%%%%%%%%%%%%%%%%%%%%%%%%%%%%%%%%%%%%%%%%%%%%
Given the distinct evolutionary histories of the two simulation cases, we examine their jet-spine topology to identify the drivers of collimation versus decollimation.

\begin{figure*}
    \centering
    \includegraphics[width=\textwidth]{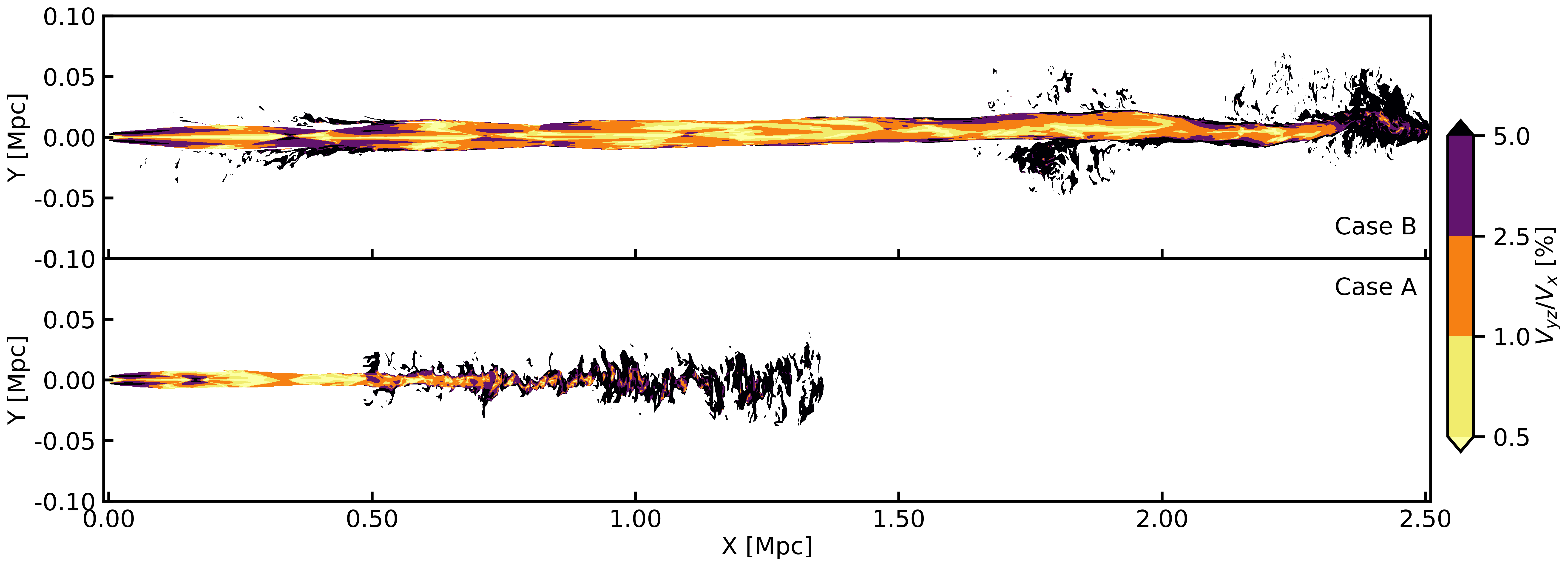}
    \caption{Distribution of the ratio of lateral velocity magnitude ($V_{yz}$) to longitudinal velocity magnitude ($V_x$) across the jet spine for our two simulation cases at their most evolved stages. The jets are injected with purely longitudinal velocity ($V_x$), while the emergence of transverse components reflects lateral expansion and collimation in the early stages, followed by wiggling motions that promote decollimation. The colourbar expresses $V_{yz}/V_x$ in percentage.}
    \label{Fig:V_ratio}
\end{figure*}

Fig.~\ref{Fig:V_ratio} shows the distribution of the ratio of lateral ($V_{yz}$) to longitudinal ($V_x$) velocity of the jet-spine, expressed in percentage (around 0.5–5\%). 
Since the jets are injected with purely longitudinal speed, the emergence of transverse components directly reflects internal jet processes. 
Even small lateral perturbations at our indicated percent level are known to seed MHD instabilities leading to a jet's motion becoming unstable \citep{Mignone2013,Mizuno2011,Mukherjee2020,Acharya2021}. 
For `\textit{Case A}', pressure imbalances at the base (jet is $\sim 2$ time over-pressured in both the cases) produce alternating expansion and reconfinement, and beyond the first few recollimation zones the induced lateral velocity begins to dominate the spine dynamics. 
This growth in lateral perturbations drives pronounced wiggling and diffusion, which reduce forward propagation efficiency and generate a broad, lobe-like structure. 
In contrast, `\textit{Case B}' maintains a more stable evolution (Fig.~\ref{Fig:V_ratio}): its first recollimation shock forms slightly farther downstream due to the higher power of the jet, and the lateral perturbations remain suppressed over most of the jet length, consistent with stabilisation by the (higher) magnetic field strength (Table~\ref{Tab:jet_params}). 
Only near the head region, where the jet head's propagation speed drops precipitously and it interacts with the turbulent cocoon, do mild undulations appear, signaling the gradual transition toward decollimation at later times.

\begin{figure*}
    \centering
    \includegraphics[width=\textwidth]{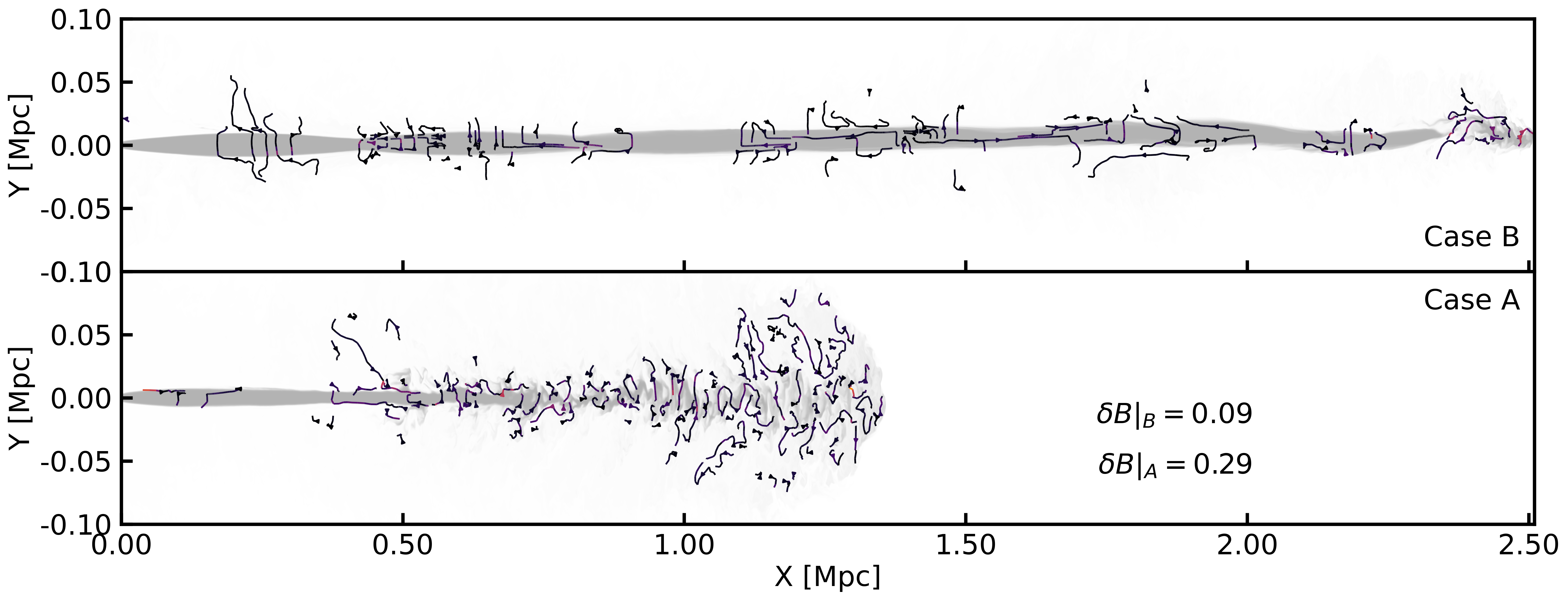}
    \caption{Magnetic field streamlines ($B_{x-y}$ presented in $x-y$ plane) of the jet spine for the two cases, showing a longer persistence of structured field in `\textit{Case~B}' compared to the earlier turbulent transition in `\textit{Case~A}'. The corresponding fractional fluctuation $\delta B$ is found to be lower in `\textit{Case~B}', consistent with its more ordered flow.}
    \label{Fig:B_lines}
\end{figure*}

In order to provide an initial qualitative assessment of the magnetic field topology, we visualize the field lines using streamlines integrated from the $B_x$ and $B_y$ components across a representative 2D planar cross-section ($x-y$ plane). 
We have also examined the extent of symmetry along the orthogonal plane; see Appendix~\ref{Sec:Alternative Planar Views of the Simulated Cases}. 
While a full 3D volumetric rendering of the magnetic field lines would offer a comprehensive topological view, we instead focus on a representative 2D analysis to circumvent the very extensive computational bottlenecks posed by the scale and memory requirement of our 3D datasets. 
Fig.~\ref{Fig:B_lines} illustrates the field topology within the jet spine, highlighting that `\textit{Case B}' preserves the toroidal nature of the injected field over a much larger spatial extent compared to `\textit{Case A}', owing to a higher injection strength and to a higher $\Gamma$ values. 
In `\textit{Case B}', later on in its spatial evolution, the toroidal field transitions into poloidal component, consistent with expectations for powerful jets where jet plasma ploughs into and remains anchored to the shear and shock surfaces, with the jet’s forward motion stretching the field in the propagation direction \citep[e.g.][also discussed in Section~\ref{Sec:Morphological and Temporal Evolution}]{Laing1981}. 
By contrast, `\textit{Case A}' shows an early onset of disordered, turbulent fields within the spine, which subsequently promote lateral diffusion of field lines, whereas the other case displays only mild undulations near the jet head. 
To quantify these trends, we further estimated the (normalized) fractional fluctuation of magnetic-field in the spine, as follows \citep{Dubey2023}:
\begin{equation}\label{Eq:Turb}
    \delta B = \frac{\sum_i (B_i - \overline{B_i})}{\sum_i B_i}
\end{equation}
computed by evaluating the difference between the local field value and the mean field within a surrounding cube of side length $5\,\mathrm{kpc}$ (which represents the typical lateral jet extent near the injection region), summed over the jet extent and subsequently normalized. 
We find that $\delta B$ is significantly lower for the collimated flow case compared to the more diffusive case (see Fig.~\ref{Fig:B_lines}), supporting the above observation.

\begin{figure*}
    \centering
    \includegraphics[width=0.45\textwidth]{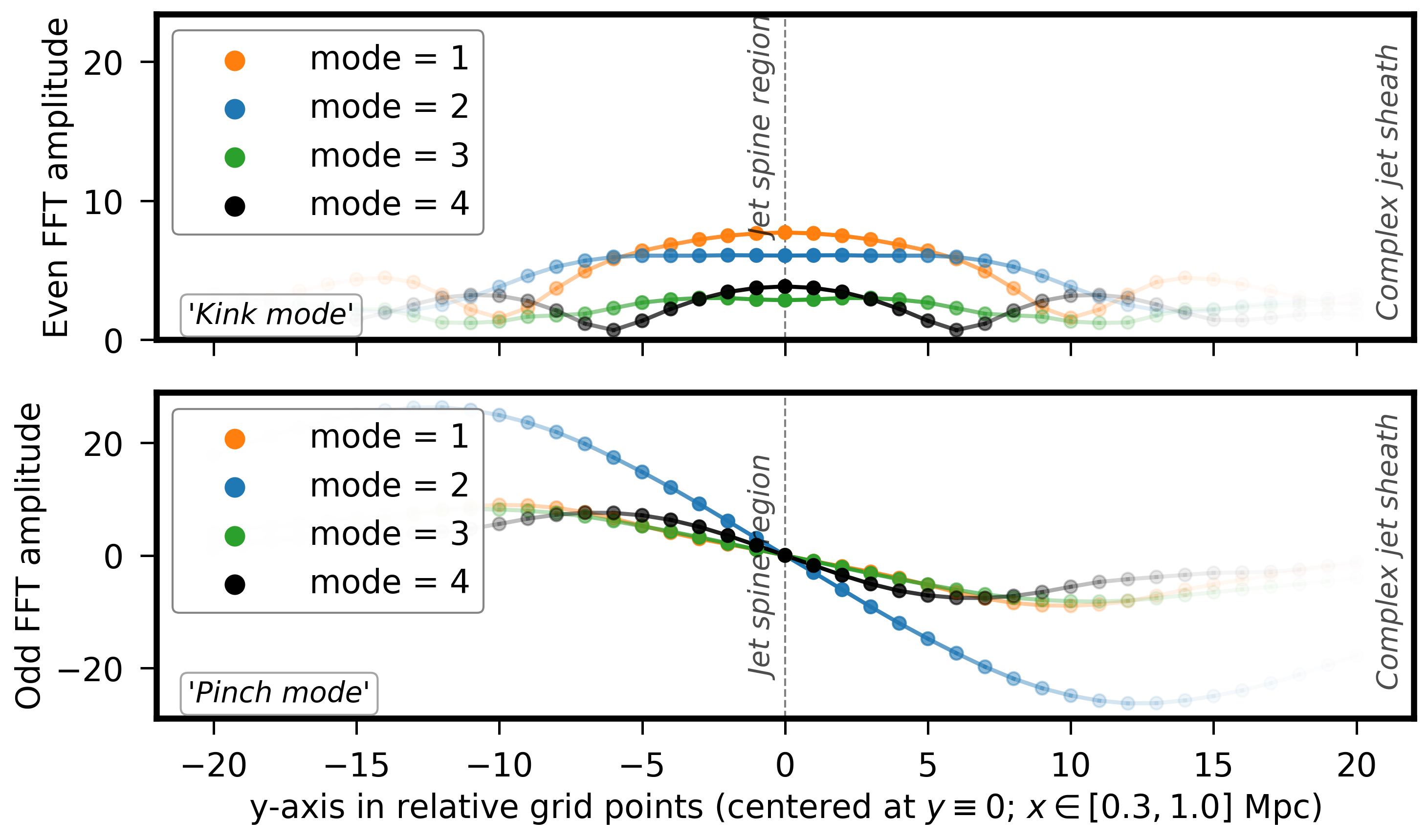}
    \hspace{0.02\textwidth}%
    \rule{0.5pt}{5cm} % vertical line: width x height
     \hspace{0.02\textwidth}
    \includegraphics[width=0.45\textwidth]{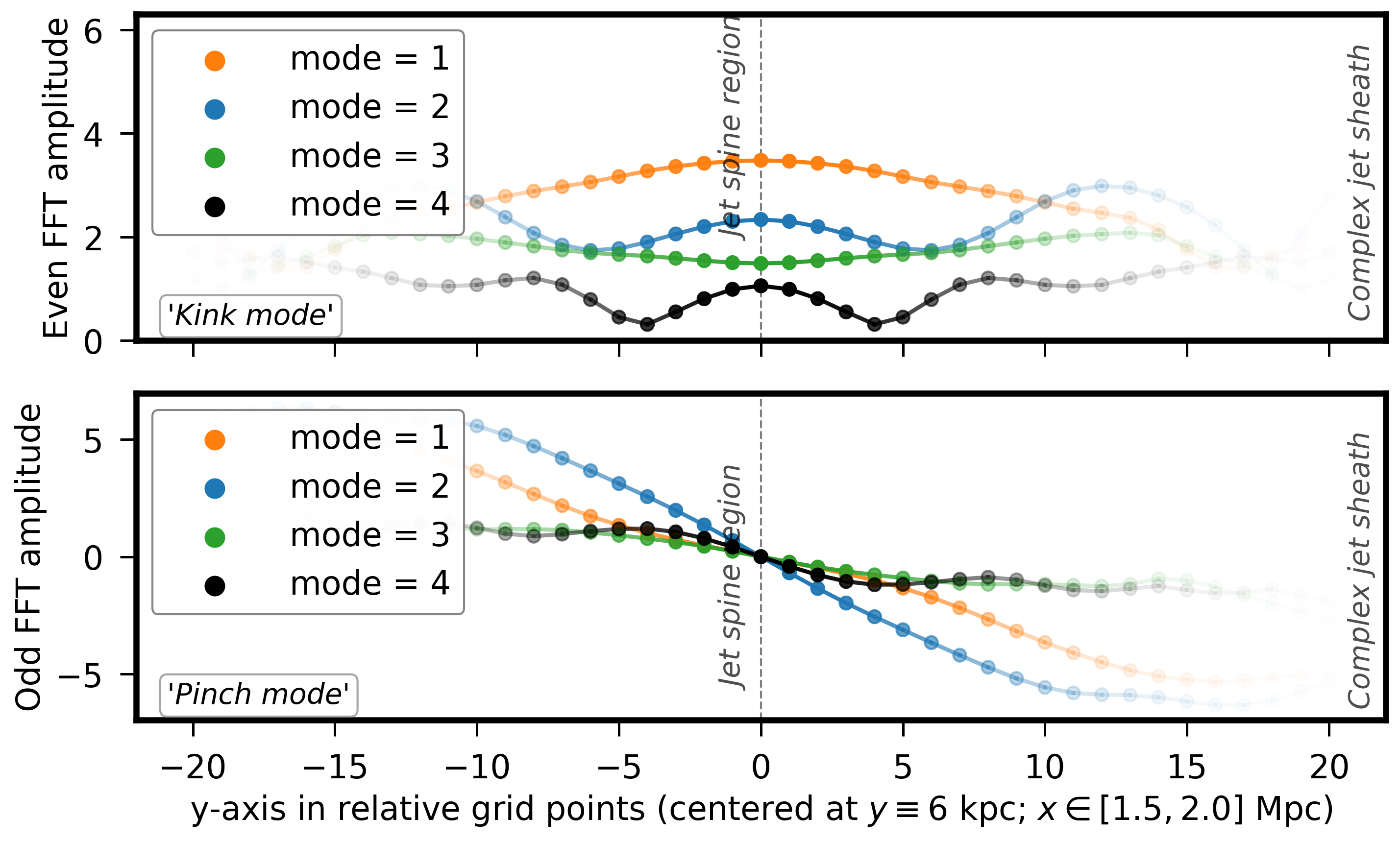}\\
    \includegraphics[width=\textwidth]{Instabilities.png}\\
    \vspace{-0.4cm}
    \includegraphics[width=0.45\textwidth]{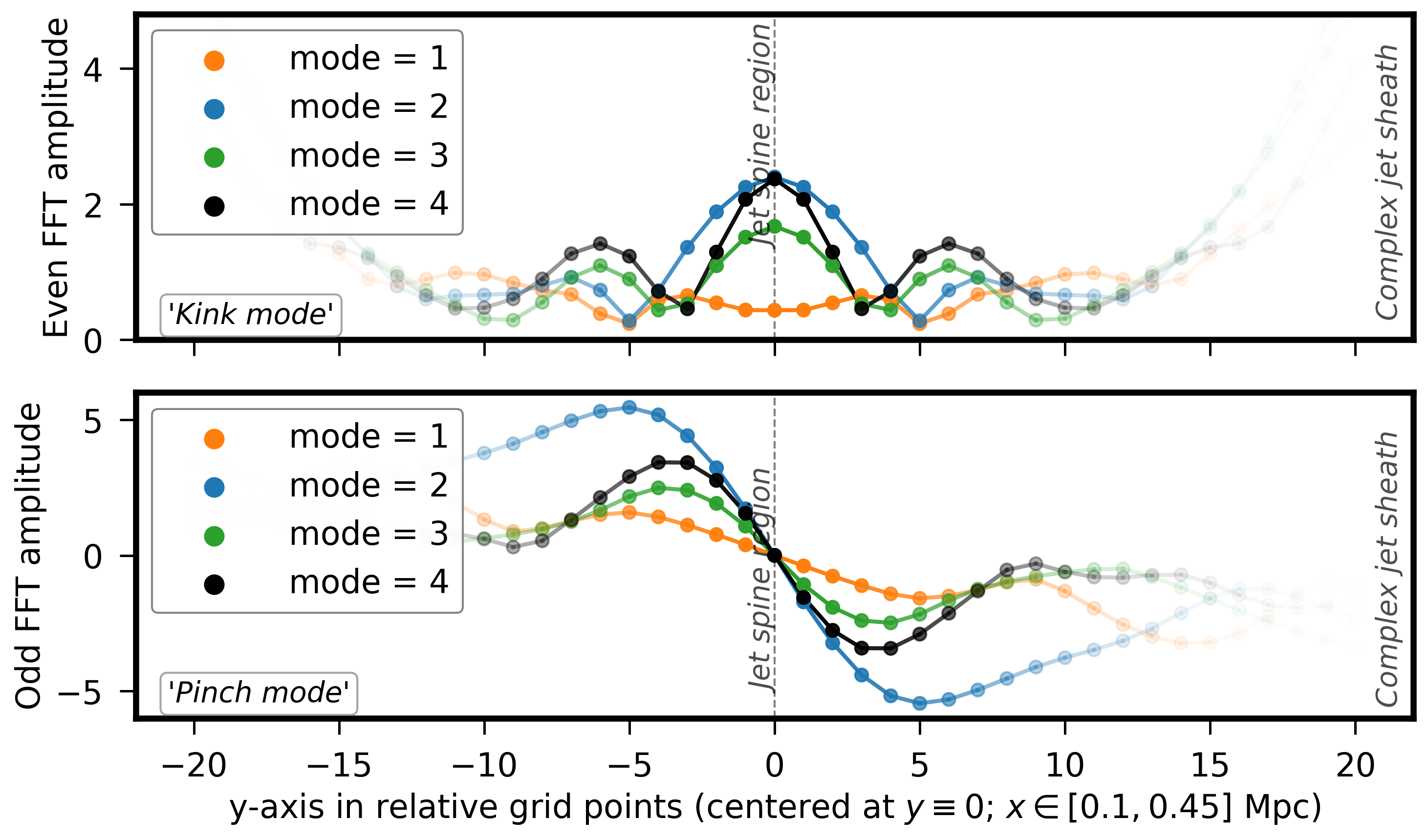}
    \hspace{0.02\textwidth}%
    \rule{0.5pt}{5cm} % vertical line: width x height
    \hspace{0.02\textwidth}
    \includegraphics[width=0.45\textwidth]{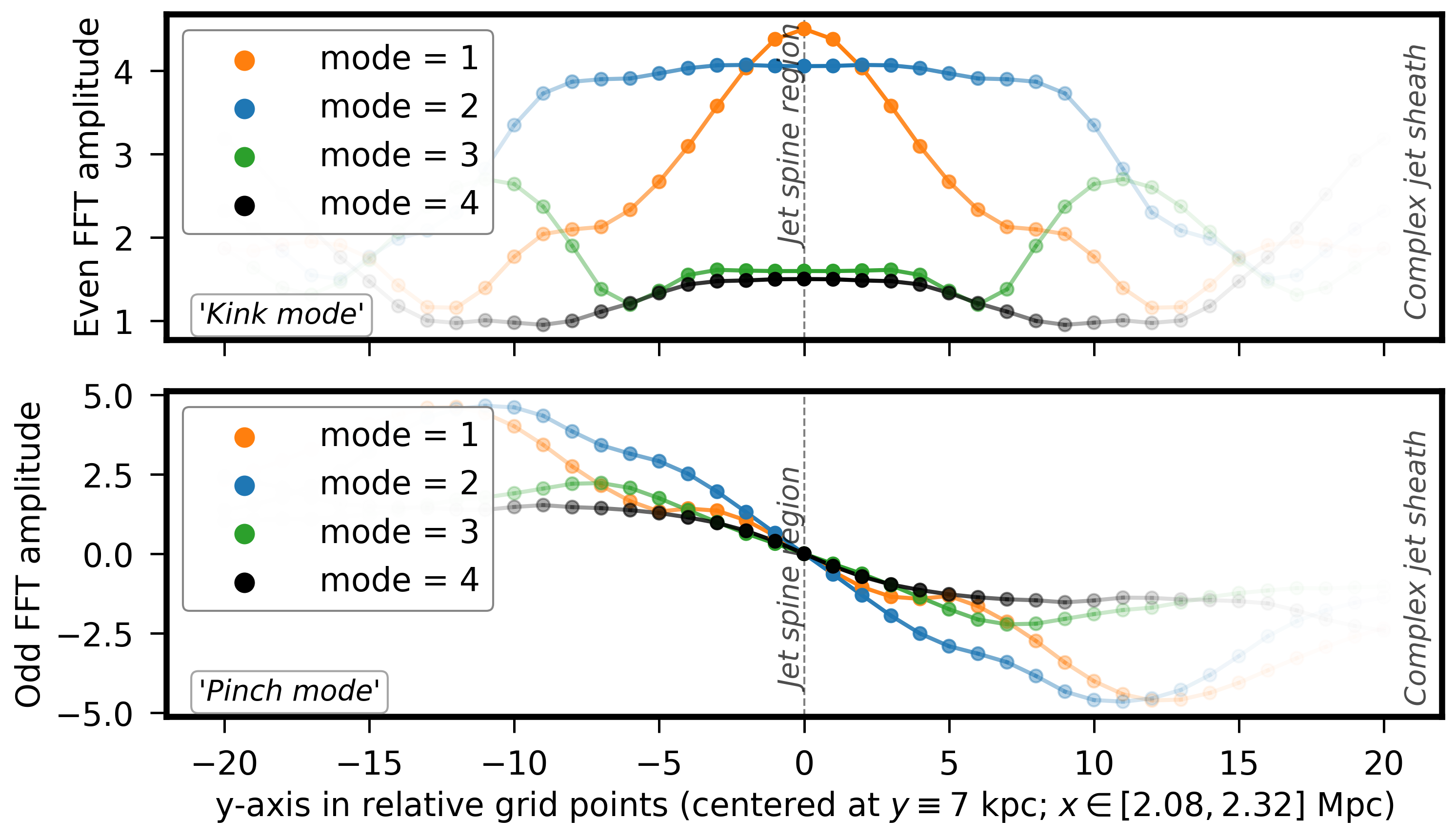}
    \caption{Fourier analysis used to quantify MHD instabilities at different locations along the jet spine for `\textit{Case~A}' and `\textit{Case~B}'. The central panels show the jet spine morphology using a colormap of the longitudinal velocity $V_x$. Black dashed rectangles mark the regions selected for the Fourier analysis: three regions for `\textit{Case~B}' and one for `\textit{Case~A}', chosen to represent a clean, approximately periodic segment of the jet spine. For each region, the transverse velocity $V_y$ is shown in (zoom-in) inset panels using a \texttt{seismic} colormap. The Fourier decomposition is performed on $V_y$, and the corresponding amplitudes of the even (kink) and odd (pinch) modes are displayed alongside each inset to identify the dominant instability modes, as inferred from their FFT amplitudes. Only the first four Fourier modes are shown, as higher-order modes contribute negligibly. The analysis is restricted to the jet spine, excluding the surrounding jet--sheath region where the velocity field is influenced by multiple overlapping processes (accordingly, the Fourier mode amplitudes smoothly diminish outside the spine).}
    \label{Fig:Instabilities}
\end{figure*}

We quantify transverse perturbations by performing Fourier
decomposition of the transverse velocity, $V_y$, along the jet spine, shown in Fig.~\ref{Fig:Instabilities} \citep[similar to,][]{Guan2014,Upreti2024}. 
For each region where the spine remains sufficiently coherent and periodic to permit a meaningful spectral decomposition via Fourier analysis (the dashed boxes in Fig.~\ref{Fig:Instabilities}), we identify the jet–spine centre $y_0$ and extract a strip extending $\pm 20$ grid cells about $y_0$. 
We then compute the Fourier transform of $V_y$ across the transverse direction. 
To separate kink--type (antisymmetric) and pinch--type (symmetric) distortions, we exploit the parity of the signal about the spine \citep{Biskamp2000}: symmetric fluctuations contribute to the even spectrum, while antisymmetric fluctuations contribute to the odd spectrum.
Let $V_y(x,y)$ denote the transverse velocity field. 
The Fourier transform across the transverse direction is:
\begin{equation}
\hat{V}(k,y) = 
\mathcal{F}\!\left[V_y(x,y)\right]
           = \sum_{x} V_y(x,y)\, e^{-jkx},
\end{equation}
where $j = \sqrt{-1}$ is the imaginary unit, and $k$ is the wave number.
For each offset $i$ around the jet spine centre $y_0$ 
(with $i = 1,\dots,20$), we form the parity–separated fields

\begin{equation}
\begin{array}{lcl}
  \hat{V}_{\mathrm{even}}(k,i)   & = & \displaystyle 
  \frac{\hat{V}(k, y_0+i) + \hat{V}(k, y_0-i)}{2},  
    \\ \noalign{\medskip}
  \hat{V}_{\mathrm{odd}}(k,i)     &=& \displaystyle 
  \frac{\hat{V}(k, y_0+i) - \hat{V}(k, y_0-i)}{2}.
\end{array}
\end{equation}
We then consider the amplitude spectra, averaged over nearby spine centers,

\begin{equation}
E_{\mathrm{even}}(k)
= \left\langle \, \hat{V}_{\mathrm{even}}(k,i) \, \right\rangle_{i,y_0},
\qquad
E_{\mathrm{odd}}(k)
= \left\langle \,  \hat{V}_{\mathrm{odd}}(k,i) \, \right\rangle_{i,y_0}.
\end{equation}
Even modes ($E_{\mathrm{even}}$) trace axisymmetric, pinch–type perturbations,
whereas odd modes ($E_{\mathrm{odd}}$) isolate lateral, kink–type distortions. Only the first four modes are considered for Fig.~\ref{Fig:Instabilities}, as higher-order terms contribute negligibly.

In `\textit{Case B}', the region near the jet base shows strong power in the odd components, with mode $k=2$ dominating, indicating a pair of pronounced pinch structures superimposed on a weaker global kink (even mode 1). 
Farther downstream, both even and odd modes contribute comparably: the spine simultaneously compresses and undergoes a gentle lateral bending, consistent with the visual deformation. 
Closer to the head, the spectrum becomes increasingly dominated by the even mode 1, reflecting the development of a large-scale kink while residual pinching persists as the jet narrows.
For `\textit{Case A}', only one interval exhibits a sufficiently periodic spine to analyze. 
There, the spectrum is dominated by odd modes, indicating progressive pinching and contraction, while weaker even modes (primarily 2 and 4) produce mild undulations but no sustained global kink. 
The FFT results show that the instability does not grow uniformly along the jet. 
The strongest peaks occur near recollimation regions, where the flow is compressed and the shear layer becomes stronger. 
Farther downstream, smaller-scale modes start to appear, indicating that large distortions gradually break up into finer structures. 
The comparison between `\textit{Cases A}' and `\textit{B}' suggests that magnetic stabilization mainly delays this nonlinear growth, rather than completely suppressing it \citep{Biskamp2000}. 
Taken together, these results indicate that well-collimated, fast jets initially favor pinch-type distortions, whereas the growth of kink modes becomes more prominent as the flow decelerates and the spine loses collimation, ultimately promoting large-scale destabilization.

%Both jets exhibit signatures of pinch and kink MHD instabilities. Pinch modes dominate during the early, highly collimated phase of fast jet propagation, while kink modes grow at later times as the flow decelerates and loses collimation, ultimately driving (or at the onset of driving) large-scale jet destabilization. ***THIS IS REDUNDANT WITH THE PREVIOUS PARAGRAPH***

%%%%%%%%%%%%%%%%%%%%%%%%%%%%%%%%%%%%%%%%%%%%%%%%%%%%%%%%%%%%%%%%%%%%%%%%%%
\section{Jet -- environment interconnection}
\label{Sec:Jet -- environment interconnection}
%%%%%%%%%%%%%%%%%%%%%%%%%%%%%%%%%%%%%%%%%%%%%%%%%%%%%%%%%%%%%%%%%%%%%%%%%%
Given the cosmological scales involved in our study, it is essential to examine the extent to which relativistic jets of such nature interact with -- and potentially reshape -- their surrounding large-scale environments, and to understand how this coupling evolves over time.

\begin{figure*}
    \centering
    \includegraphics[width=\textwidth]{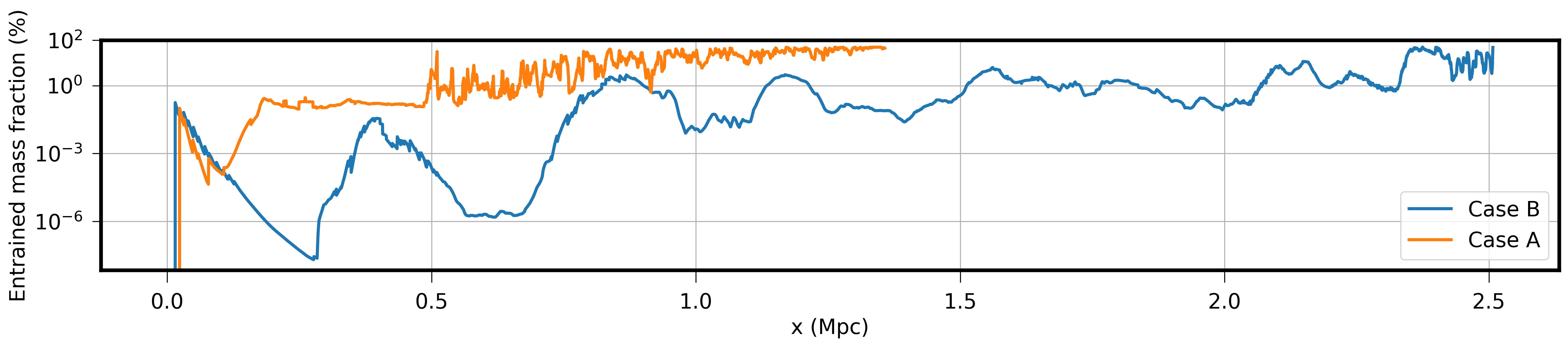}
    \caption{Percentage of ambient medium mass entrained into the jet spine. A clear global increase in entrainment toward the jet head is observed, with notably higher entrainment in `\textit{Case A}' (at 34.3 Myr) than in `\textit{Case B}' (at 15 Myr), highlighting the key role of entrainment in jet destabilization.}
    \label{Fig:Entrainment}
\end{figure*}

In this regard, the entrained mass fraction shown in Fig.~\ref{Fig:Entrainment} measures the contribution of ambient material ($M_{\rm amb}$) to the total mass ($M_{\rm amb} + M_{\rm jet}$) contained within the jet spine (region with jet tracer $\mathcal{T}_j \geq 0.5$). 
Close to the inlet the fraction remains extremely small, indicating that the freshly injected jet is initially only weakly contaminated, although it does not vanish entirely, consistent with early shear-driven mixing and interaction with the cocoon backflow. 
Further downstream the profile develops pronounced oscillations that trace the sequence of recollimation shocks, with the amplitude of these variations staying significant as the flow propagates. 
Superimposed on these local shock–compression cycles is a clear global rise of the entrained fraction, reflecting cumulative ambient mass loading of the jet spine. 
Toward the jet head the fraction approaches order unity, showing that the spine becomes extensively loaded with the ambient material and is likely to experience significant deceleration (see Fig.~\ref{Fig:V-L}). 
The effect is markedly stronger in `\textit{Case A}', where the weaker, more decollimated jet mixes more efficiently with the cocoon, whereas `\textit{Case B}' remains better collimated and protected, although it still exhibits substantial mass loading near the head. 
In `\textit{Case B}', the poloidal component of the magnetic field is stronger, and the overall field magnitude is higher (Fig.~\ref{Fig:1D_variations_AB}). 
This enhanced magnetic support stabilizes the jet spine and, in turn, suppresses shear-driven instabilities and limits entrainment \citep{Biskamp1998,Wang2023,Rossi2024}. 
Since entrainment itself is a key driver of jet decollimation \citep[e.g.][]{Massaglia2016,Abolmasov2023}, the reduced mixing effectively forms a positive feedback loop with the stronger field, and together this loop keeps the flow better protected during its propagation compared to `\textit{Case A}'.

\begin{figure*}
    \centering
    \includegraphics[width=\textwidth]{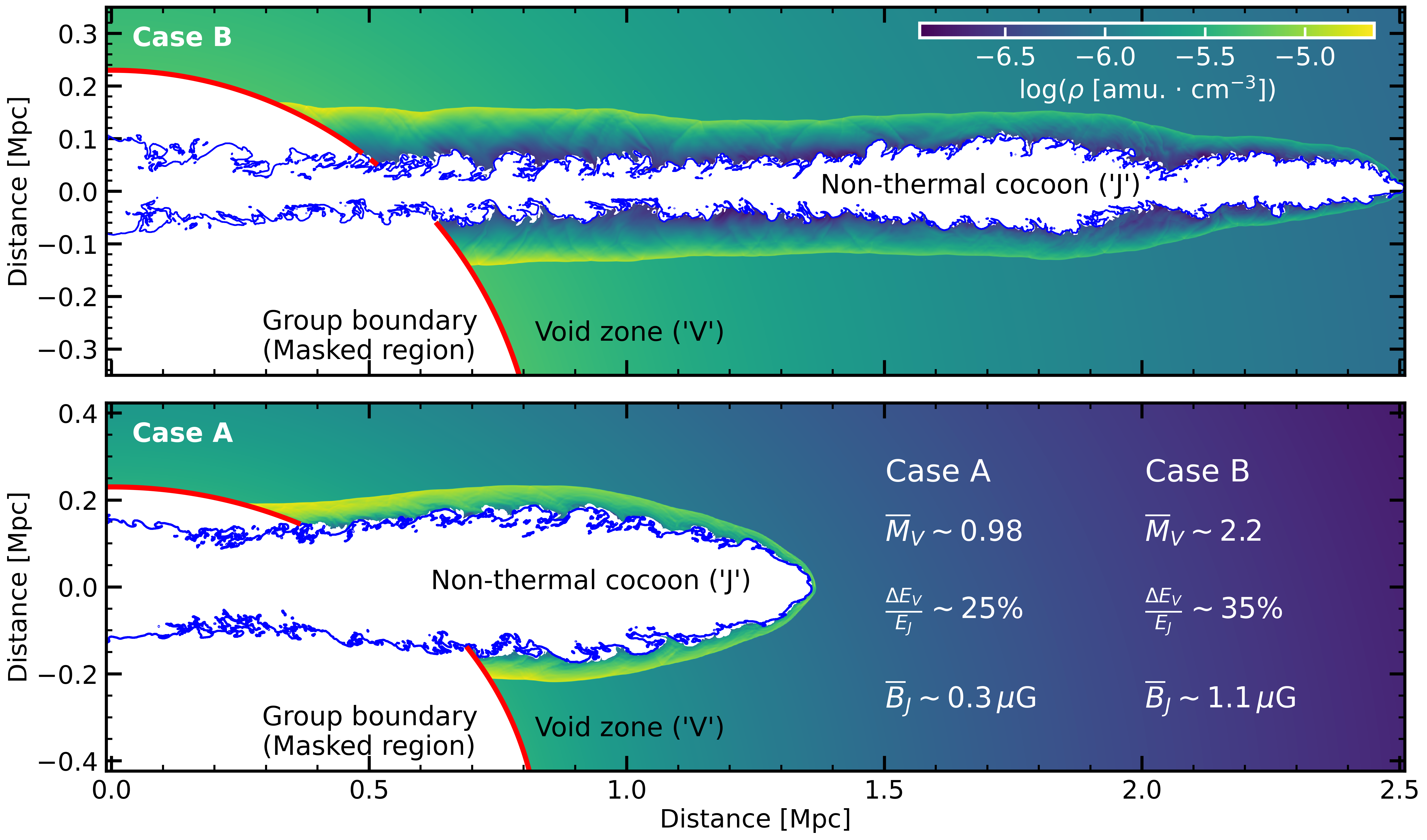}
    \caption{Density slices for `\textit{Case A}' (bottom) and `\textit{Case B}' (top), illustrating the transport of matter, energy, and magnetic fields by AGN jets launched from a galaxy group and propagating into the surrounding void. The galaxy group region, defined within the virial radius, is excluded from all diagnostics to isolate jet feedback on the void. The non-thermal jet cocoon (`J') and the cocoon-driven shock propagating into the void (`V') are indicated. We quantify the efficiency of energy transport into the void by measuring the accumulated energy gain of the void material (the accumulated energy increment $\Delta E_V$ at time $t$ relative to the initial epoch $t = 0$) relative to the jet energy deposited beyond the group boundary ($E_J$ at $t$). Time $t$ is 34.3 Myr for `\textit{Case A}' and 15 Myr for `\textit{Case B}'. The sonic Mach number ($\overline{M}_V$) of the cocoon-driven shock in the void and the magnetic field ($\overline{B}_J$) transported by the non-thermal cocoon are measured to characterise the dynamical and magnetisation impact of the jet. The colour scale shows the logarithmic gas density.}
    \label{Fig:Void_eedback}
\end{figure*}

We further investigate the impact of powerful radio jets propagating from the outskirts of a galaxy group into an underdense large-scale environment, focusing on the transfer of energy and magnetic fields into the surrounding void (Fig.~\ref{Fig:Void_eedback}). 
In our simulations, the jet is launched from a location approximately 600 kpc above the group center and subsequently propagates beyond the group virial radius (Fig.~\ref{Fig:setup}), which we adopt as a working boundary between the group environment and the void. So, the void region is defined as the volume beyond the galaxy group’s virial radius of 812 kpc (from center ($0,0,0$); Fig.~\ref{Fig:setup},~\ref{Fig:Void_eedback}). 
While the transition between these cosmological structures is intrinsically gradual, the current choice allows an approximate, physically motivated separation of the two regions.
The galaxy group interior is therefore masked from all diagnostics, and all reported quantities are computed exclusively within the void region.

We quantify the efficiency of energy transport into the void by measuring the energy gain of the void material (the accumulated energy increment $\Delta E_V$ at time $t$ relative to the initial epoch $t = 0$) relative to the jet energy deposited beyond the group boundary ($E_J$ at $t$). 
Time $t$ is 34.3 Myr for `\textit{Case A}' and 15 Myr for `\textit{Case B}'. 
For the low-power, more dissipative jet (`\textit{Case A}'), we find that approximately 25\% of the jet energy is transferred to the void, whereas the higher-power jet (`\textit{Case B}') deposits roughly 35\% of its energy into the same region. 
We note that while measuring $\Delta E_V$, the jet cocoon zone (jet tracer $\mathcal{T}_{j} \geq 10^{-7}$) is excluded (Fig.~\ref{Fig:Void_eedback}), ensuring that the measured energy increase reflects genuine energy transfer to the void.

\begin{figure*}
    \centering
    \includegraphics[width=\textwidth]{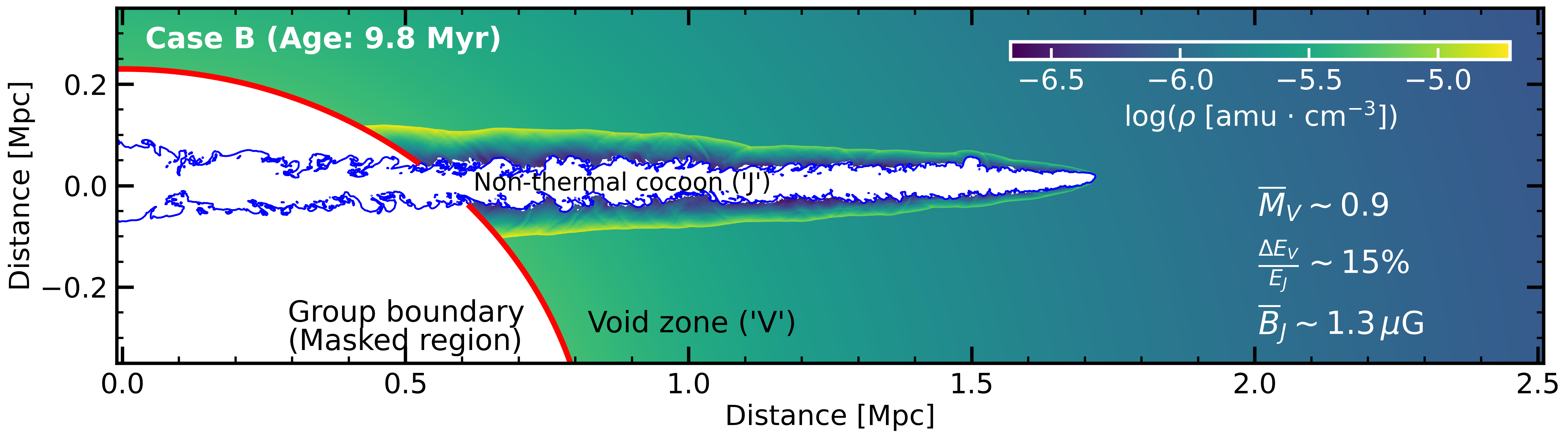}
    \caption{Same as Fig.~\ref{Fig:Void_eedback}, but shown for `\textit{Case B}' at an earlier evolutionary epoch in order to isolate the influence of evolutionary age on the jet--environment coupling discussed in Fig.~\ref{Fig:Void_eedback}, particularly in the context of the comparatively higher jet injection power of `\textit{Case B}'. The sonic Mach number associated with the shock transporting energy into the void region, together with the average magnetic field strength carried by the jet material into the void, are evaluated at this evolutionary stage (9.8 Myr) for `\textit{Case B}'.}
    \label{Fig:Void_eedback_CaseB_earlyTime}
\end{figure*}

To better understand the physical mechanism responsible for this energy transfer, we examine the strength of the shock outside the non-thermal cocoon (we note again the exclusion of regions with $\mathcal{T}_{j} \geq 10^{-7}$ and the galaxy group within its $812$ kpc virial radius; Fig.~\ref{Fig:Void_eedback}).  
The sonic Mach number is estimated from the ratio between the density-weighted velocity \citep{Mignone2013} of the shocked void material and the characteristic sound speed of the ambient medium. For the adopted thermodynamic setup in our simulations, the corresponding adiabatic sound speed is $\sim 540$ km s$^{-1}$ \citep[see, e.g.,][]{Mendygral2012}. We note that this value follows directly from the simplified equilibrium density--pressure configuration adopted for the ambient medium (Section~\ref{Sec:Ambient medium configuration}). In comparison, galaxy clusters, owing to their substantially hotter intracluster media, typically exhibit higher sound speeds \citep[e.g.,][]{Oneill2010,Mendygral2011}.
We estimate the sonic Mach number of the shocked void medium ($\overline{M}_V$) and find a marked difference between the two cases. 
In `\textit{Case A}', the affected void medium's expansion is transonic, with a mean Mach number, $\overline{M}_V \sim 1$, indicating that energy dissipation occurs primarily through weak shocks and sound waves. 
In contrast, `\textit{Case B}' exhibits a substantially stronger shock, with $\overline{M}_V \sim 2.2$, enabling more efficient conversion of jet kinetic energy into thermal and kinetic energy of the ambient void medium. 
The higher void energy fraction measured in `\textit{Case B}' can therefore be naturally attributed to the stronger shock-driven coupling between the jet-driven cocoon and its environment. 
The above estimates are derived from snapshots corresponding to different evolutionary stages of the jets, selected from the final stages of the respective simulations. To provide an additional perspective on the jet--environment coupling, Fig.~\ref{Fig:Void_eedback_CaseB_earlyTime} also presents the same quantities for `\textit{Case B}' at an earlier evolutionary epoch, owing to its higher injection power (in comparison to `\textit{Case A}'). Ideally, a direct comparison between the two simulation cases would be performed at epochs corresponding to identical cumulative injected jet energies. However, owing to the discrete temporal cadence adopted for saving simulation outputs in these computationally demanding runs (a single snapshot occupies $\sim  1.6$ terabytes of storage space), no snapshot was available that simultaneously satisfied this condition for both cases. The comparatively lower value of $\overline{M}_V$ and the reduced fraction of energy transferred to the ambient medium at this earlier stage (in Fig.~\ref{Fig:Void_eedback_CaseB_earlyTime}) primarily reflects the ballistic nature of the jet propagation during the initial phase of evolution, restricting lateral growth of the cocoon, where a larger fraction of the injected energy remains confined within the rapidly advancing linear jet channel (aside from the naturally shorter duration available for jet--environment interaction).

In addition to energy transport, we assess the efficiency of magnetic field injection into the void. Since the jet cocoon has not yet dispersed into the surrounding void medium (Fig.~\ref{Fig:Case_evolution}), our estimate of magnetic-field transport to the void accounts only for the field advected beyond the group boundary by the jet. The magnetic energy remains largely confined within the cocoon, and quantifying the timescale over which it mixes into the void is deferred to future work.
By computing the jet mean magnetic field strength ($\overline{B}_J$) within the void region (Fig.~\ref{Fig:Void_eedback}), we find that `\textit{Case A}' deposits a mean field of approximately $0.3 \, \mu$G, while `\textit{Case B}' transports a notably larger field of $1.1 \, \mu$G. 
These values are non-negligible when compared to typical magnetic field strengths expected in large-scale, underdense environments \citep[which are typically an order of magnitude lower;][]{Stuardi2020}, suggesting that powerful radio jets may play an important role in magnetising cosmic voids. 
The substantially higher magnetic field strength in `\textit{Case B}' further reflects the enhanced dynamical impact of higher-power (and higher magnetised) jets on the surrounding medium. We note that the values of $\overline{B}_J$ for `\textit{Case B}', when compared between Figs.\ref{Fig:Void_eedback} and \ref{Fig:Void_eedback_CaseB_earlyTime} at different evolutionary stages, indicate that the average magnetic field strength within the cocoon decreases as the cocoon volume expands. This behaviour is consistent with the progressive dilution and redistribution of magnetic energy during the cocoon evolution \citep[e.g.,][]{Giri2026_JetDir}. 

Aside from the the global energetics, Fig.~\ref{Fig:Void_eedback} reveals several morphological features that provide further insight into the nature of jet–environment coupling at the group–void interface. 
The expansion of the non-thermal cocoon and its surrounding shock is slightly asymmetric, reflecting the presence of ambient density gradients \citep[anticipated, given the jet's injection at the group boundary;][]{Giri2025_Emission}. 
The topology of the shocked ambient material (outside the jet cocoon) differs significantly between the two cases: the lower-power jet (`\textit{Case A}') remains relatively laterally confined, consistent with a more dissipative, momentum-dominated regime, whereas the higher-power jet (`\textit{Case B}') produces a broader, laterally inflated shocked ambient region, characteristic of an energy-driven expansion. 
In `\textit{Case B}', the shocked shell compresses void gas to densities approaching those found near the group outskirts, demonstrating that powerful jets can locally restructure the thermodynamic state of under-dense environments. 
The cocoon-driven shock in this case also remains coherent over several hundred kiloparsecs, indicating efficient long-range jet-driven disturbances within the void.

We emphasise that the present analysis is intended as a preliminary demonstration of jet feedback on large-scale cosmic structures. 
A more detailed treatment—examining the long-term dissipation of injected energy, its spatial redistribution within the void, and the cumulative impact on void thermodynamics—will require dedicated future studies. 
Furthermore, the response of the galaxy group itself, which is expected to be highly sensitive to AGN feedback, remains an important avenue for investigation \citep{Cavaliere2008,Randall2015}. 
Nevertheless, our results demonstrate that extreme-scale AGN jets can transport substantial energy and magnetic flux beyond their host environments, implying that their influence on the cosmic web may be significant and should not be neglected \citep{Oei2024_7Mpc}.

\section{Conclusions}\label{Sec:Conclusions}
%%%%%%%%%%%%%%%%%%%%%%%%%%%%%%%%%%%%%%%%%%%%%%%%%%%%

The discovery of GRGs with projected jet lengths reaching 5–7 Mpc has pushed classical models of AGN jet propagation to their limits, raising fundamental questions about how relativistic jets can remain collimated and radiatively visible over possibly required Gyr timescales.
Those inferred ages of the most extreme systems challenge radiative cooling constraints and demand a deeper understanding of how jets survive magnetohydrodynamical instabilities, as well as whether sustained jet launching from the central black hole can persist over such long durations. 
Motivated by these challenges, we performed targeted numerical simulations of two jet configurations that probe distinct dynamical regimes of thrust and collimation, while remaining representative of powerful FR II-type systems. 
`\textit{Case A}' represents a standard powered FR II–type jet with moderate Lorentz factor, larger injection radius, and weaker magnetic support, while `\textit{Case B}' represents a higher-thrust, more tightly collimated jet with a larger Lorentz factor, and stronger magnetic field characterizing a somewhat more powerful jet. 
By injecting jets at the boundary of a poor galaxy group, consistent with the environments of extreme GRGs, we investigate the physical mechanisms that enable sustained, multi-megaparsec jet propagation.

Our simulations demonstrate that the ability of oppositely directed pairs of jets to reach extreme ($\sim 5$ Mpc) GRG scales is governed primarily by how efficiently the jet spine preserves its Lorentz factor, collimation, and magnetic coherence during propagation. 
In `\textit{Case B}', the combination of higher injection Lorentz factor ($\Gamma = 10$), reduced jet radius, slightly enhanced jet-to-ambient density contrast, and stronger magnetic field produces a fast, tightly collimated spine that efficiently delivers thrust to the jet head. 
Early over-expansion into an under-pressured environment drives strong, quasi-periodic recollimation shocks, which—rather than dissipating the flow—promote rarefaction-driven acceleration between shocks, thereby allowing the spine to reach $\Gamma \sim 20$ and driving jet propagation faster within the rarefraction-and-compression zones. 
The strengthened poloidal field, generated by jet flow induced shear and compression at recollimation shocks and at jet boundaries, suppresses transverse MHD instabilities and damps lateral perturbations, leading to ballistic-like propagation with limited entrainment. 
As a result, the jet advances rapidly ($\sim 5$ Mpc in bi-directional flow extent in $\sim 15$ Myr), demonstrating that even the most extreme GRGs indeed can be produced within the standard $10^7$ yr timescales.

By contrast, `\textit{Case A}', despite being a powerful FR II-type jet, occupies a parameter regime in which collimation cannot be sustained beyond the initial couple of recollimation zones. 
The lower injection Lorentz factor ($\Gamma = 7$), larger jet injection radius, and weaker magnetic field result in progressively damped recollimation shocks, enhanced coupling between the spine and the turbulent cocoon, and increased mass loading. 
As the jet interacts more strongly with backflows and ambient material, the Lorentz factor profile flattens, the spine transitions from a coherent flow to a diffuse structure, and thrust delivery to the head becomes inefficient. 
This promotes lateral expansion and lobe formation but still can naturally produce $\sim 3$ Mpc-scale GRGs (in bi-directional total flow extent), albeit over somewhat longer jet lifetimes. 

During these active evolutionary phases, the total injected energy into the ambient environment differs significantly between the two cases, with `\textit{Case B}' (in 15 Myr) corresponding to $2.3 \times 10^{61}$ erg and `\textit{Case A}' to $8.1 \times 10^{60}$ erg (in 34.3 Myr).

Our analysis identifies clear signatures of both pinch and kink MHD instabilities developing along the jet spine. 
Pinch modes dominate during the early (spatially), strongly collimated phase of jet propagation, consistent with compression-driven oscillations associated with recollimation shocks. 
At later stages (downstream of jet flow), as the jet decelerates and interacts more strongly with the cocoon, kink modes become increasingly prominent, producing lateral distortions. 
These results indicate that the growth of MHD instabilities is an intrinsic aspect of jet evolution on multi-megaparsec scales and is unlikely to be entirely avoided. 
Instead, the existence of extreme GRGs implies that specific regimes of injected jet speed, magnetic structure, and collimation can delay instability growth sufficiently to permit exceptionally extended propagation. 
Identifying these regimes across a broader parameter space holds the potential to address how common such extreme jet propagation can be.

A well-collimated jet naturally separates into a fast, high-$\Gamma$ relativistic spine and a slower, dissipative jet head, where strong interaction with the ambient medium regulates the advance speed and generates backflowing material. 
In `\textit{Case B}', the overall jet structure grows at $\sim 0.5 c$, nearly an order of magnitude faster than the speeds typically assumed in observational age estimates for extreme GRGs, and reaches even higher values at earlier evolutionary stages. 
In `\textit{Case A}', the head decelerates more rapidly to $\sim 0.05c$, yet still exceeds commonly adopted values by a factor of a few, with early-time propagation speeds remaining large when compared to standard assumptions. 

Our simulations demonstrate that extreme-scale AGN jets can provide efficient feedback beyond their host galaxy groups, transferring substantial energy and magnetic flux into surrounding voids. 
The higher-power, better-collimated jet (`\textit{Case B}') couples more effectively to the ambient medium, driving a stronger cocoon shock, depositing a larger fraction of its energy ($\sim 35\%$), and transporting a magnetization of $\mu$G-level strengths. 
In contrast, the lower-power jet (`\textit{Case A}') exhibits weaker, largely transonic coupling and reduced magnetic transport, reflecting its more dissipative and entrainment-dominated evolution. 
These results indicate that powerful radio jets may play a non-negligible role in heating and magnetizing under-dense regions of the cosmic web.

Finally, our proxy emission maps indicate that synchrotron radiation is spatially localized in regions of enhanced compression and magnetic field amplification—primarily near the first recollimation shocks at the jet base and at the jet head interaction zone—reflecting the physical conditions likely responsible for the observed emission patterns in extreme GRGs.

\subsection*{Present limitation and future extension}

Our simulated jets exhibit a linear advance speed somewhat higher than typically inferred in conventional radio galaxy studies. While higher, this value emerges self-consistently from our physically motivated initial conditions and assumed environment. Given the limited observational constraints on growth speeds for extreme giant radio galaxies, firm conclusions remain premature. We note that environmental complexity may influence propagation. The present model adopts a smooth ambient medium; incorporating turbulence or a more structured density profile (e.g., a double-beta model including a galactic potential) could introduce additional resistance \citep{Mendygral2012,Dutta2024}, particularly during the early expansion phase. These refinements are planned for future work. We note here that while recent radio surveys have significantly increased the number of known GRGs \citep{Mostert2024}, they have also revealed a broader diversity of radio morphologies \citep{Yang2019,Bhukta2022,Horton2025,Lochner2025}. It remains plausible that extreme-scale sources still constitute a relatively small subset of the overall radio galaxy population, in which case comparatively rapid large-scale growth may not be unexpected. Further observational constraints and statistical samples will be essential for placing stronger limits on jet advance speeds at extreme scales.

In the present study, we vary a limited set of physically motivated parameters to identify a viable baseline model for extreme-scale jet propagation. This serves as an initial step towards a full parameter exploration.
In future work, we will perform a systematic parameter study to isolate the role of individual physical quantities in controlling jet stability and large-scale collimation. This will include controlled variations of jet thrust, Lorentz factor, magnetization, and magnetic-field topology, as well as nozzle properties such as cylindrical versus finite opening-angle injection. The goal is to quantitatively determine which parameter regimes allow jets to remain stable and collimated to multi-megaparsec scales.

\section*{Acknowledgements}

We thank the anonymous referee for their constructive suggestions, which have helped improve and enrich the manuscript.  We acknowledge the Istituto Nazionale di AstroFisica (INAF) for awarding this project access and computational time to the LEONARDO supercomputer, owned by the EuroHPC Joint Undertaking, hosted by CINECA (Italy) and the LEONARDO consortium.
Scientific results presented in this work were obtained with the aid of the gPLUTO code, for which the authors acknowledge funding from the European High Performance Computing Joint Undertaking (JU) and Belgium, Czech Republic, France, Germany, Greece, Italy, Norway, and Spain under grant agreement No. 101093441 (SPACE). The authors thank the SPACE consortium for making the simulation code publicly available. 
%%%%%%%%%%%%%%%%%%%%%%%%%%%%%%%%%%%%%%%%%%%%%%%%%%
\section*{Data Availability}

The simulation source files and the resulting data from this study are available from the corresponding authors upon reasonable request.



\bibliographystyle{mnras}
\bibliography{sample} % if your bibtex file is called example.bib

%\begin{thebibliography}{99}
%\end{thebibliography}

%%%%%%%%%%%%%%%%%%%%%%%%%%%%%%%%%%%%%%%%%%%%%%%%%%

%%%%%%%%%%%%%%%%% APPENDICES %%%%%%%%%%%%%%%%%%%%%

\appendix

\section{Alternative Planar Views of the Simulated Cases}\label{Sec:Alternative Planar Views of the Simulated Cases}

\begin{figure*}
    \centering

    % -------- Left panel --------
    \begin{overpic}[
      width=0.45\textwidth,
      trim={0cm 0cm 0.3cm 1cm},
      clip
    ]{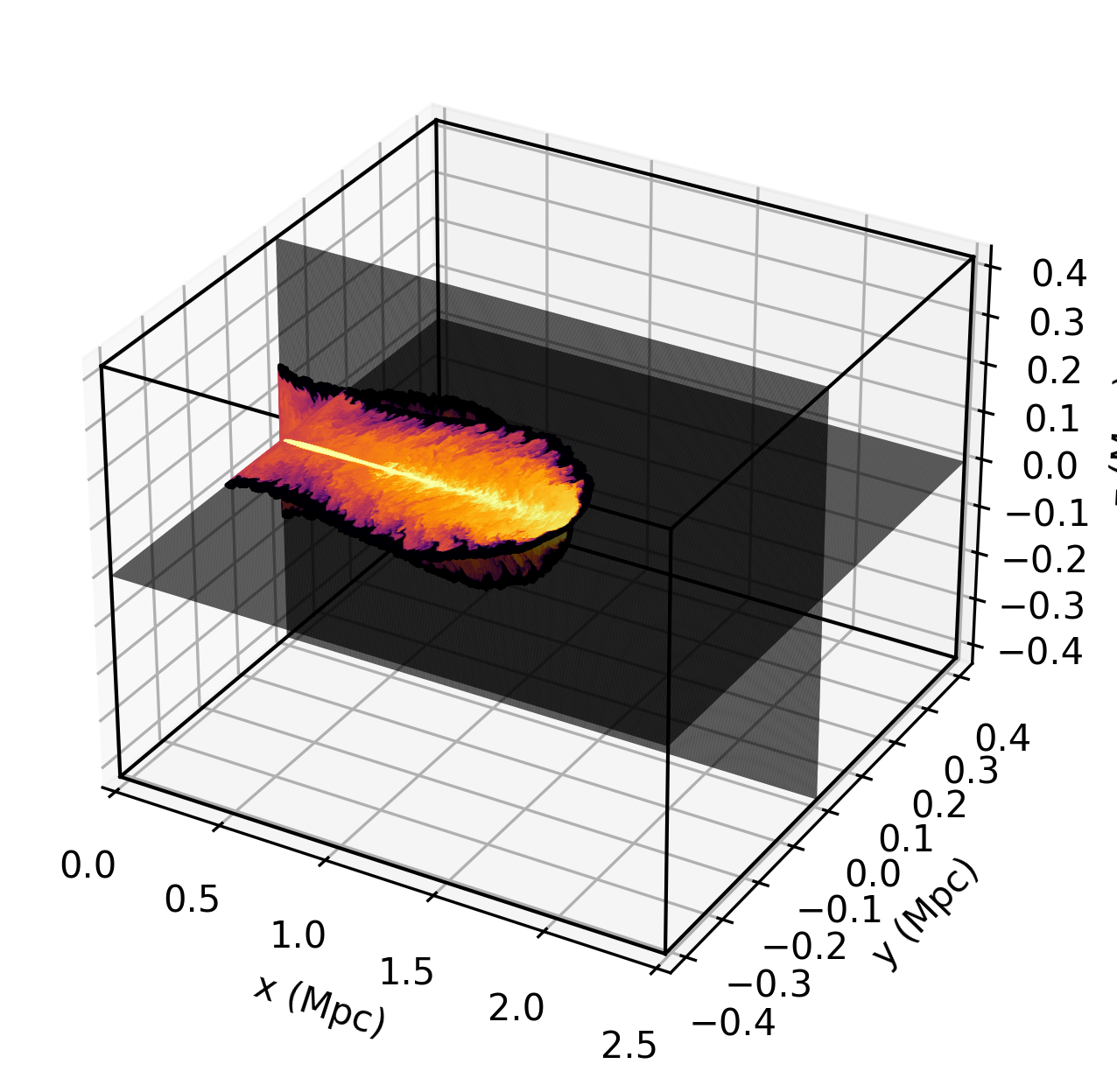}
    \put(101,50){\rotatebox{90}{$z\,(\mathrm{Mpc})$}}
    \end{overpic}
    \hspace{0.2cm}
    % -------- Right panel --------
    \begin{overpic}[
      width=0.45\textwidth,
      trim={0cm 0cm 0.3cm 1cm},
      clip
    ]{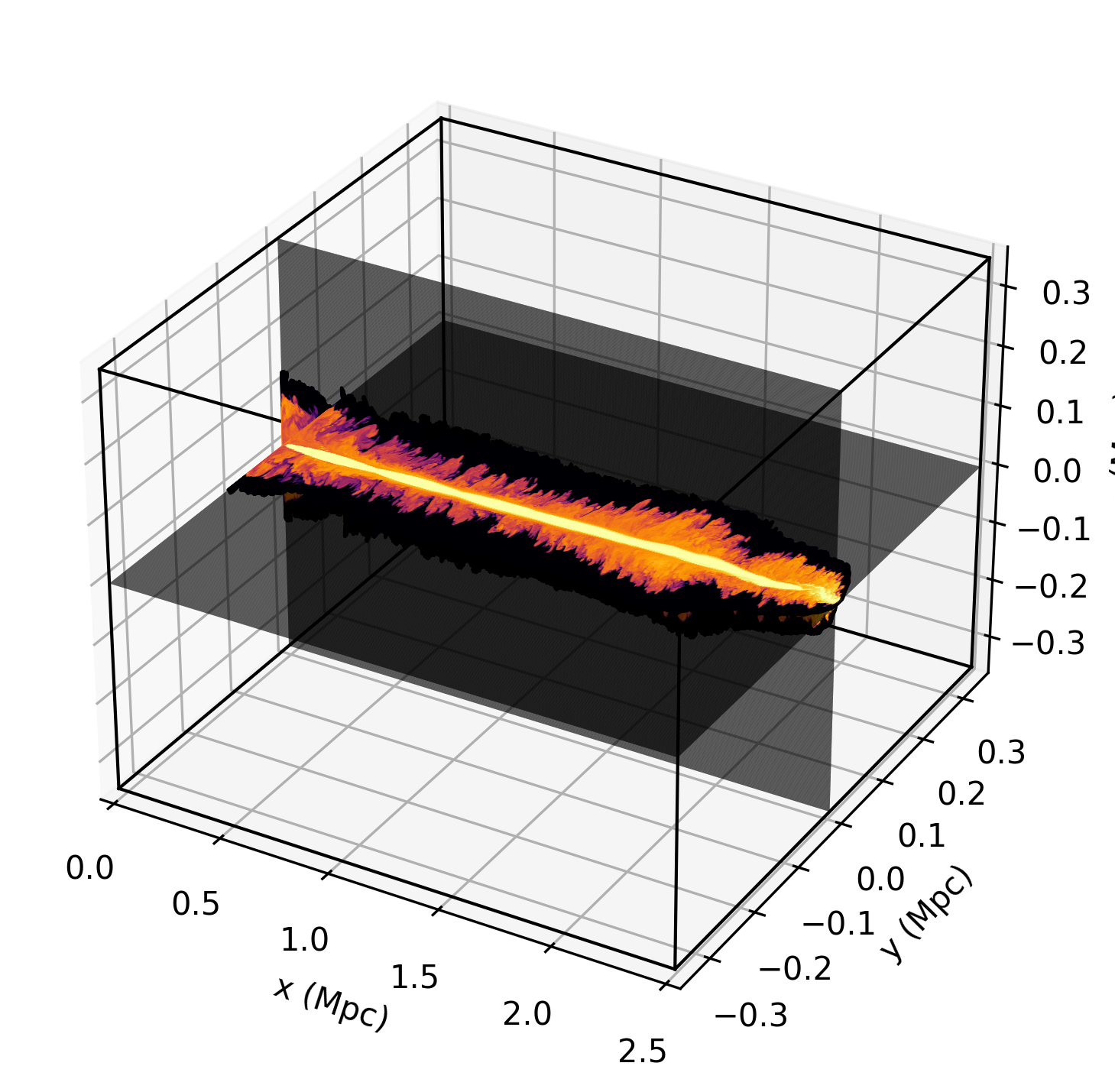}
      \put(101,50){\rotatebox{90}{$z\,(\mathrm{Mpc})$}}
    \end{overpic}

    \caption{
    Three-slice 3D visualization of the evolved jet structures for `\textit{Case A}' (\textit{left}) and `\textit{Case B} (\textit{right}) showing the tracer values at 34.3 Myr and 15 Myr, respectively. Due to the prohibitive data volume required for a full 3D volume rendering, the jet morphology is illustrated using orthogonal planar cuts: an $x-y$ slice and an $x-z$ slice embedded within the 3D computational domain. This provides a complementary view to Fig.~\ref{Fig:Case_evolution} and highlights the differing small-scale activity around the jet spine and jet--environment shear boundaries, which cannot be reproduced in 2D simulations.
    }
    \label{Fig:tracer_xzPlane}
\end{figure*}

To better envisage the three-dimensional picture of the jet flows in our simulated cases (Table~\ref{Tab:jet_params}), Fig.~\ref{Fig:tracer_xzPlane} presents $x-y$ and $x-z$ slices plotted as the three-slice 3D visualization manner of the jet tracer values, serving as complementary views to Fig.~\ref{Fig:Case_evolution}. Given the substantial size of the data sets produced by our simulations (a single double-precision data file output occupies 1.6 terabytes), we adopt this planar-slice visualization as a practical alternative to full three-dimensional, kernel-based volume rendering. This approach preserves essential three-dimensional physical insights while remaining analytically tractable. Qualitatively, the tracer morphology in the $x-z$ plane broadly resembles that seen in the $x-y$ slices shown (also) in the main discussion of this work (see, Fig.~\ref{Fig:Case_evolution}), indicating that the planar analyses performed throughout the paper are not expected to alter our overall conclusions. 

However, this apparent similarity does not imply that the system can be adequately captured using two-dimensional simulations. Although the large-scale structure appears broadly symmetric, the jet beam exhibits clear plane-dependent differences in its fine structure, including variations in spine width, wiggling patterns, and local distortions. Such features are inherently three-dimensional and play a critical role in governing mass entrainment, the growth of MHD instabilities, magnetic-field amplification, and ultimately the jet’s propagation efficiency.

%%%%%%%%%%%%%%%%%%%%%%%%%%%%%%%%%%%%%%%%%%%%%%%
\section{Influence of the host galaxy and the global intragroup medium}\label{Sec:Influence of the host galaxy and the global intragroup medium}

As discussed in Section~\ref{Sec:Ambient medium configuration}, GRGs are preferentially found in relatively low-density large-scale environments, such as poor galaxy groups and cosmic filaments \citep{Stuardi2020}, while generally avoiding rich X-ray luminous galaxy groups and cluster environments \citep{Andernach2021,Simonte2024}. The most extended systems, particularly those exceeding projected linear sizes of $\sim 3$ Mpc, are also observed to preferentially reside in comparatively underdense environments \citep{Oei2022_5Mpc,Oei2024_7Mpc,Sankhyayan2024}. Furthermore, only 0.34\% of BCGs in dense environments are known to host GRGs, while amongst the GRG population only 8.4\% is associated with BCG systems \citep{Dabhade2020_LOTSS}. More broadly, studies of radio-loud AGN populations in galaxy groups by \citet{Kolokythas2018,Kolokythas2019} indicate that the central regions of groups are predominantly occupied by radio-quiet systems. Even among systems originating from the central dominant galaxy (cD galaxy), the typical sizes of such RGs generally remain below $\sim 1$ Mpc \citep{Giacintucci2007}. Furthermore, even within the GRG population originating from cD galaxies, the median projected linear size of jetted systems is reported to be only $\sim 0.92$ Mpc \citep{Dabhade2020_VLASS}. Extending this to more massive environments, \citet{Kale2015} showed that BCGs residing in dynamically unrelaxed clusters are comparatively less likely to host radio-loud AGN activity, suggesting an additional environmental constraint on the formation and long-term growth of large RGs.

These observational trends motivate the simulation setup adopted in Table~\ref{Tab:jet_params}, where the jets are injected from the outskirts of the formulated group environment rather than from the dense central region \citep[cf.,][]{Oei2023_spiralHost}. Such a configuration was intentionally chosen to investigate whether powerful jets propagating through comparatively low-density group outskirts can naturally evolve into multi Mpc-scale radio structures similar to the recently observed GRG population. Nevertheless, in the current context, it is also relevant to examine how the jet evolution changes if the same jetted system instead is hosted by a centrally located AGN associated with the first-ranked cD-type galaxy or BCG of the group. This scenario may be particularly relevant for the comparatively small number of reported GRGs ($\lesssim 2$ Mpc) that appear to be associated with BCG-like host systems \citep{Subrahmanyan2008,Giacintucci2011}. 
   
\begin{table*}
\centering
\caption{Simulation parameters for additional validation runs corresponding to Table~\ref{Tab:jet_params}.}
\label{Tab:jet_params_Appendix}
\begin{tabular}{lcccccccc}
\hline\hline
Simulation label & $\Gamma$ & $r_{\mathrm{j}}$ [kpc] & $\rho_{\mathrm{j}}/\rho_{0}$ & $\sigma$ & $Q_{\mathrm{j}}$ [erg s$^{-1}$] & Description & Domain [kpc$^3$] & Ambient Medium\\
& &  & & &  &  & & [amu/cc] [kpc]\\
\hline
`\textit{Case B: Gr\_center}' & $10$ & $2.0$ & $5 \times 10^{-5}$ & $0.10$ & $ 2.4 \times 10^{46}$ & Jet flow from group's center & $2520 \times 700 \times 700$ & $\rho_0: 10^{-3}$\\
& &  & & &  &  & (jet flow along $x$) & $a = b = c: 33$\\
`\textit{Case A/B: El\_galaxy}' & $7$ & $0.2$ & $10^{-5}$ & $0.01$ & $ 2.3 \times 10^{46}$ & Jet flow inside a galaxy & $12 \times 16 \times 12$ & $\rho_0: 1.0$\\
& &  & & &  &  & (jet flow along $y$) & $a: 1,\,\, b: 2.7,\,\, c: 1.4$\\
\hline
\end{tabular}
\begin{tablenotes}
\small
\item \textbf{Notes.} {Top entry corresponds to a jet identical to `\textit{Case B}' in Table~\ref{Tab:jet_params}, but injected from the center of the galaxy group profile. The bottom entry presents a simulation with a jet of comparable power and parameter regime to that in Table~\ref{Tab:jet_params}, performed as a demonstrative study of jet propagation within a galactic environment on substantially smaller spatial scales. The ambient medium parameters listed at the end of the table correspond to Eq.~\ref{Eq:Ambient_prof_Appendixx}. The numerical resolution in both simulations follows the same criterion adopted throughout this work, i.e., 5 cells per jet radius. The definitions of the remaining columns are identical to those in Table~\ref{Tab:jet_params}}.

\end{tablenotes}
\end{table*}

To explore the above possibility, we performed an additional simulation in which the same jet as `\textit{Case B}' in Table~\ref{Tab:jet_params} is injected from the center of the group atmosphere formulated in Section~\ref{Sec:Ambient medium configuration}, thereby allowing the outflow to interact with the full group-scale potential. The ambient density profile is written in Cartesian coordinates as,
%%%%%%%%%%%%%%%%%%%%%
\begin{equation}\label{Eq:Ambient_prof_Appendixx}
\rho(x,y,z) = \rho_0 \left[1 + \left(\frac{x}{a}\right)^2 + \left(\frac{y}{b}\right)^2 + \left(\frac{z}{c}\right)^2 \right]^{-\frac{3}{2}\beta},
\end{equation}
%%%%%%%%%%%%%%
where the parameters $a$, $b$, and $c$ represent the core radius ($\equiv r_c$, and $r_0 = 0$ in Eq.~\ref{Eq:Ambient_prof}). All other jet and ambient parameters are kept same as those adopted in `\textit{Case B}'. This additional setup therefore represents a jet launched from the center of the same galaxy group environment, and is hereafter referred to as `\textit{Case B: Gr\_center}'. The relevant simulation parameters are summarized in Table~\ref{Tab:jet_params_Appendix}.

\begin{figure*}
    \centering
    \includegraphics[width=\textwidth]{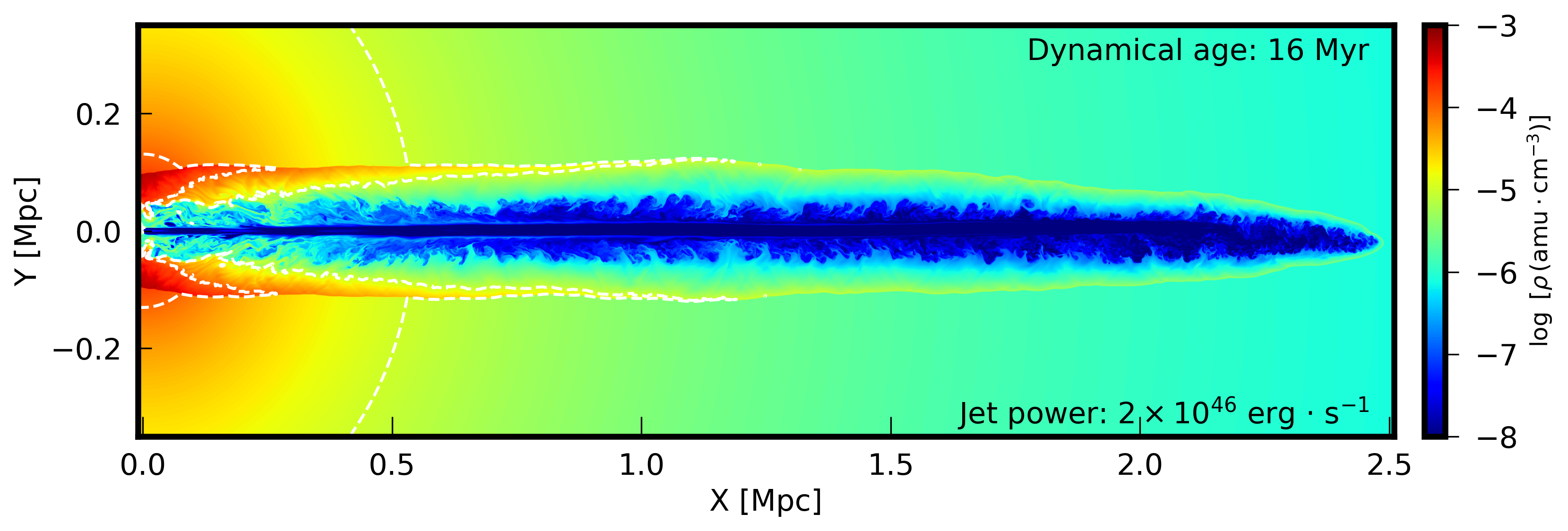}
    \caption{$x-y$ slice ($z = 0$) of the density distribution for the `\textit{Case B: Gr\_center}' simulation, where the jet is injected from the center of the galaxy group profile (contours at log $\rho \equiv -5, -4$). The resultant morphology demonstrates a spatio-temporal evolution broadly similar to that of `\textit{Case B}', reinforcing the modest influence of poor galaxy  group environments on the propagation of high-power jets. The well-collimated large-scale structure strongly indicates the potential for even larger-scale jet evolution within plausible radio-galaxy lifetimes.}
    \label{Fig:Gr_Centre}
\end{figure*}

\begin{figure}
    \centering
   \includegraphics[width=\columnwidth]{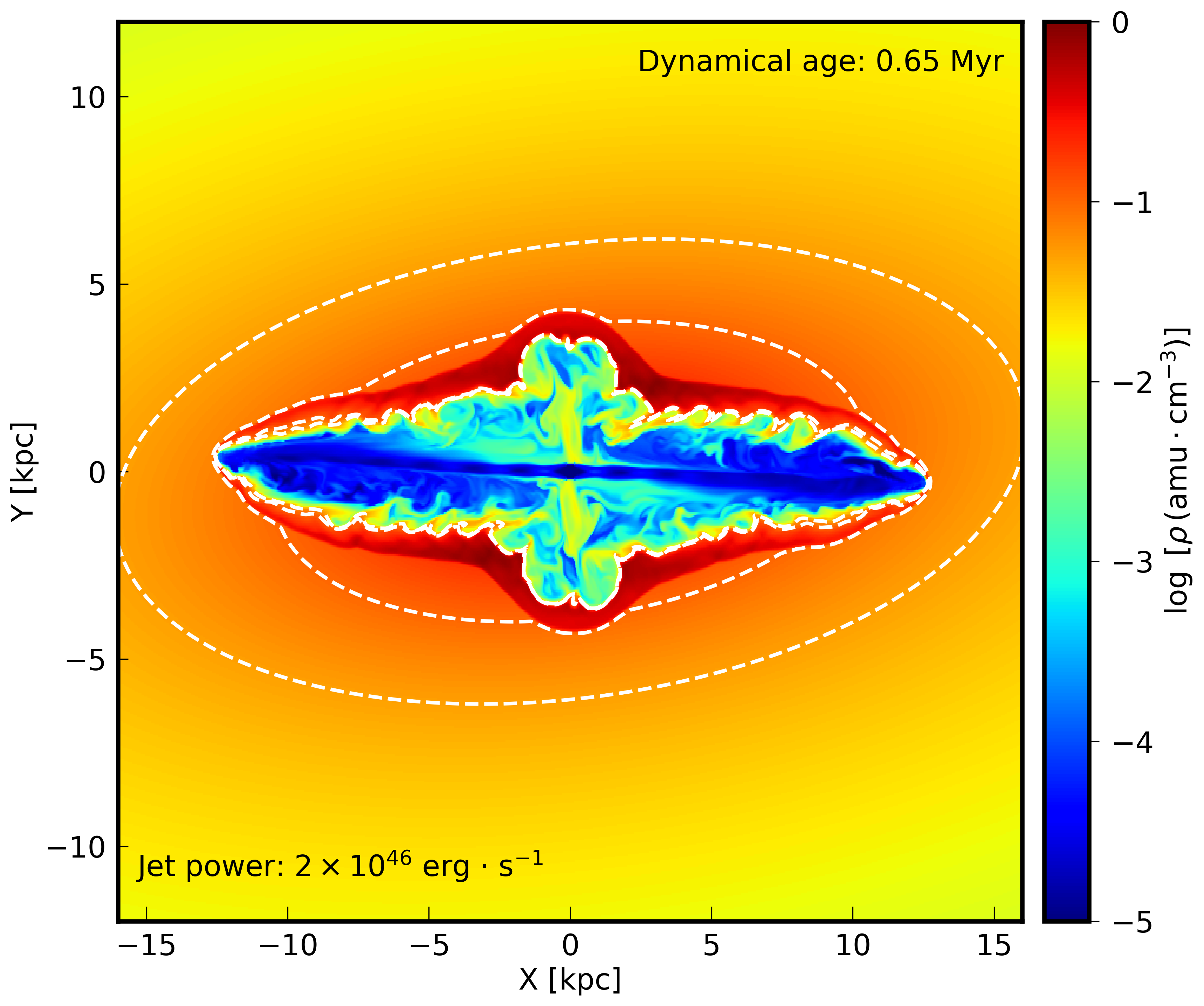}
    \caption{Demonstration of jet propagation through a galactic environment, presented via $x-y$ density slice ($z = 0$) for the `\textit{Case A/B: El\_galaxy}' simulation. The contours correspond to $\log \rho \equiv -1.3, -1$. The figure highlights the rapid propagation of the powerful jet even along the major axis of the elliptical galaxy, representing the direction of maximum environmental hindrance to the jet flow.}
    \label{Fig:El_galaxy}
\end{figure}

We present the jet–ambient interaction for the `\textit{Case B: Gr\_center}' simulation at a dynamical time of 16 Myr through a density slice in Fig.~\ref{Fig:Gr_Centre}. The overall morphology shows a close similarity with `\textit{Case B}', in which the jet propagates through the outskirts of the same group environment.
The time evolution between `\textit{Case B}' and `\textit{Case B: Gr\_centre}' is found to be negligibly different, indicating that for a sufficiently powerful jet (here, $2.4 \times 10^{46}$ erg s$^{-1}$), the early-stage propagation is only weakly influenced by the deeper group potential \citep[similar to, e.g.,][]{Oei2024_LuminousGRGs}. Once the jet exits the denser regions of the poor group environment, its evolution becomes comparable to that in a lower-density, void-like medium (as in `\textit{Case B}'). The presence of a narrow cocoon, an arrow-head–like morphology, and a well-collimated jet structure suggests continued propagation and potential growth into a larger-scale radio system. A detailed investigation of such late-time evolution and environmental dependence will be explored in a future study.

It is also useful to provide a preliminary assessment of jet propagation within a galactic medium, where the interstellar medium density is substantially higher than the mean group environment. However, the spatial scales required to self-consistently simulate jet launching from galactic cores and follow their evolution up to the large scales considered in this work (i.e., extreme GRGs) are computationally prohibitive, particularly for CPU-based setups and remain extremely challenging for GPU-accelerated codes due to the wide dynamical ranges involved. 

As a first-order investigation, we therefore perform a separate simulation in which a galaxy-scale potential is included, and a jet is injected with a power equivalent to that of `\textit{Case B}'. Owing to the significantly different spatial scales and ambient density conditions, certain parameters are adjusted accordingly; however, we ensure that they remain as consistent as possible with those adopted in `\textit{Case A}' or `\textit{B}' (Table~\ref{Tab:jet_params}).
The corresponding ambient density profile, based on Eq.~\ref{Eq:Ambient_prof_Appendixx}, together with the jet parameters chosen to reproduce the same jet power as in `\textit{Case B}' (using Eq.~\ref{Eq:JetPower}), are summarized in Table~\ref{Tab:jet_params_Appendix}. These choices are consistent with previous studies of jet propagation in galactic environments \citep[e.g.,][]{Rossi2017,Giri2022_XRG}. This additional simulation is referred to as `\textit{Case A/B: El\_galaxy}', as the adopted setup results in an elliptical galaxy–like environment with a triaxial core, where the major axis is aligned along the $y-$direction and the jet is injected and propagated along this axis. The choice of an elliptical galaxy configuration is consistent with observational inferences for GRG host systems \citep{Giri2026_JetDir}, including even the most extended sources \citep{Willis1974,Oei2022_5Mpc,Oei2024_7Mpc}, where host galaxy morphologies are often found to be relevant to `massive' ellipticals (galaxy mass $\gtrsim 10^{11.5} M_{\odot}$).

The resultant evolution at a dynamical time of $0.65$ Myr is presented in Fig.~\ref{Fig:El_galaxy}, where the jet already reaches spatial extents comparable to the typical extent of emission-line nebulae detectable in jetted elliptical galaxies.
The rapid forward propagation of the jet is particularly noticeable, indicating that the galactic potential only weakly hinders the jet flow at such high jet powers (here, $2.3 \times 10^{46}$ erg s$^{-1}$). This behaviour therefore provides additional support for our choice of directly launching the jets into the group-scale environment in the primary simulations, especially also considering the substantial computational challenges associated with resolving jet propagation self-consistently across galactic-group-to-void scales.
This result is also qualitatively consistent with the discussions presented in \citet{Potter2015} and \citet{Blandford2019}, where it was argued that jets with mechanical powers exceeding $\sim 5 \times 10^{45}\ \mathrm{erg\ s^{-1}}$ are expected to be only minimally influenced by the host galaxy environment. In this context, \citet{Giri2026_JetDir} demonstrated that jets with powers of order $\sim 3 \times 10^{44}\ \mathrm{erg\ s^{-1}}$ can still be significantly affected by the galactic potential, whereas the jets considered in the present work are substantially more powerful.
We further emphasize that the present numerical exercise should be regarded only as a preliminary investigation of the problem (jet propagation across spatio-temporal and multi-phase domain). A more rigorous exploration of jet evolution across galactic and group environments would require several additional dedicated simulations, similar in spirit to studies such as those of \citet{Yates-Jones2021} and \citet{Giri2026_JetDir}, which we have planned as a natural extension of the current work.

%%%%%%%%%%%%%%%%%%%%%%%%%%%%%%%%%%%%%%%%%%%%%%%%%%

% Don't change these lines
\bsp	% typesetting comment
\label{lastpage}
\end{document}